\documentclass[10pt]{article}

\usepackage{amsfonts,amssymb,amsmath,mathrsfs,latexsym,bbm,dsfont,amsthm}
\usepackage{color}
\usepackage{graphicx,footnote}
\usepackage{pstricks,pstricks-add}
\usepackage{pst-plot}
\usepackage{enumerate}
\usepackage[scriptsize,nooneline,hang]{caption}
\usepackage[hang,nooneline,scriptsize]{subfigure}
\usepackage[ansinew]{inputenc}

\definecolor{lightgrey}{rgb}{0.95,0.95,0.95}\definecolor{grey}{gray}{0.7}

\newpsobject{showgrid}{psgrid}{subgriddiv=1,griddots=10,gridlabels=6pt}

\begin{document}

\title{Analytic treatment of geodesics in five-dimensional Myers-Perry space--times}

\author{Valeria Kagramanova and Stephan Reimers \\ \\
Institut f\"ur Physik, Universit\"at Oldenburg,
D--26111 Oldenburg, Germany
}

\maketitle

\begin{abstract}

We present the complete set of analytical solutions of the geodesic equation in the five-dimensional Myers-Perry space-time with equal rotation parameter in terms of the Weierstra{\ss}' elliptic and Weierstra{\ss}' zeta and sigma functions. We study the underlying polynomials in the polar and radial equations which depend on the parameters of the metric and conserved quantities of a test particle and characterize the motion by their zeros. We exemplify the efficiency of the analytical method on the orbits of test particles.

\end{abstract}

\section{Introduction}

The Myers-Perry~\cite{MP86} space-time is a higher dimensional generalization of the Kerr space-time which describes rotating black holes in more than four dimensions. Interest in higher dimensional gravity is a result of the search for the theory of quantum gravity which should describe the Universe at its beginning. With (Super)String and M- theory as candidates, it finds applications in the AdS/CFT correspondence. Another interesting question is the creation of black holes at the LHC which from an energetical point of view is only in higher dimensions possible (associated with a change of the fundamental value of the Planck mass in higher dimensions). A review of the higher dimensional black hole solutions in vacuum and in supergravity theory is presented in~\cite{EmpRe08,Em08} and a discussion of the physical properties of higher dimensional black holes is addressed in~\cite{Kanti04}. General Kerr-de Sitter and Kerr-NUT-AdS metrics in all dimensions are presented in~\cite{GLPP05,ChLP06}. A very good motivation to study the higher dimensional black holes is proposed in~\cite{KoKoZhi10}.

In the present work we study the motion of test particles in the 5D Myers-Perry space-times. For simplicity of the analysis we have chosen equal values of the rotation parameters. The case with unequal rotation parameters is currently under study~\cite{KaRe11}. The motion in the Kerr space-time is presented in~\cite{Chandrasekhar83, Oneil}. Some special cases of the motion in 5D Myers-Perry space-times are studied in~\cite{FroSto03}, where it was shown that there are no stable circular orbits in the equatorial place, and in~\cite{GooFro08} scattering and capture of particles by 5 dimensional rotating black holes are investigated. Geodesic stability of circular orbits in singly-spinning $d$-dimensional Myers-Perry space-times is studied in~\cite{Cardoso09}. Properties of the motion of charged test particles around a black hole immersed into a magnetic field are studied in~\cite{Kaya07}. Separability of the Hamilton-Jacobi equations of motion in the Kerr-de Sitter and Myers-Perry space-times in all dimensions are investigated in~\cite{VaStePa05,VaStePa05_2} and in~\cite{Frolov1,Frolov2,Frolov3,Frolov4,Frolov5}. Analytical solutions of the geodesic equations in the Kerr-de Sitter space-time in 4D are studied in~\cite{HKKL09,HLKK10}. A general analytical solution of the geodesic equations in terms of the hyperelliptic theta- and sigma- functions in the Myers-Perry space-time with one rotation parameter is presented in~\cite{EHKKL2011}.

In this paper after the introduction of the equations of motion in Sec.~\ref{sec:EOM}, we discuss in Sec.~\ref{theta-r-features} the properties of test particle motion on the basis of the polar and radial equations and derive in Sec.~\ref{sec:solution} the complete set of analytic solutions of the geodesic equation in the Myers-Perry space--time in 5 dimensions for equal values of the rotation parameters in terms of the Weierstra{\ss}' elliptic, zeta- and sigma- functions. In Sections~\ref{sec:orbits} and~\ref{sec:observables} we plot the orbits for chosen sets of parameter values and discuss observable quantities.

\section{The geodesic equation}\label{sec:EOM}

The metric of a five-dimensional rotating black hole in Boyer-Lindquist coordinates has the form~\cite{MP86,GLPP05,GK09,FroSto03} 
\begin{equation}
ds^2 = -dt^2  + \rho^2\left( \frac{du^2}{4\Delta} + d\vartheta^2 \right) + \alpha\sin^2\vartheta d\varphi^2 + \beta\cos^2\vartheta d\psi^2 + \frac{r_0^2}{\rho^2}\left( dt + a\sin^2\vartheta d\varphi + b\cos^2\vartheta d\psi \right)^2 \, , \label{metrikMP}
\end{equation}
where $\rho^2=u+a^2\cos^2\vartheta + b^2\sin^2\vartheta$ and $\alpha = u + a^2$, $\beta = u + b^2$, $\Delta=\alpha\beta - r_0^2 u$. Here we have introduced the coordinate $u$ such that: $u=r^2$ and $r$ is a radial coordinate. The parameter $r_0^2$ is proportional to the mass of the black hole.

\begin{figure}[htbp]
\centering
   \includegraphics[width=0.3\textwidth]{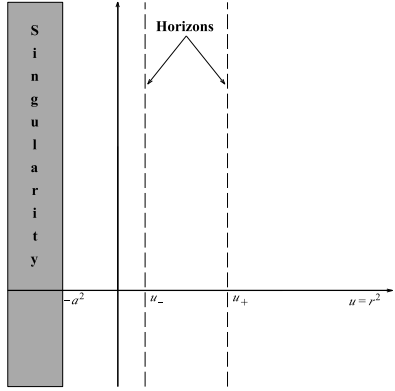}
\caption{Position of the singularity and the horizons in the Myers-Perry black hole with equal rotation parameter $a=b$.}
\label{fig:sing}
 \end{figure}

The metric is singular at $\Delta=0$ and $\rho^2=0$. There are also angular coodinate singularities at $\vartheta=0$ and $\vartheta=\frac{\pi}{2}$. Calculating the Kretschmann scalar one sees that the surface $\rho^2=0$ is a scalar curvature singularity~\cite{GK09} since the Kretschmann scalar diverges at $\rho^2=0$. Thus, for $a=b$ the singularity is located at $u=-a^2$~\cite{GK09} as shown in Fig.\ref{fig:sing}. The zeros of $\Delta$ define the horizons which for $a=b$ yield
\begin{align}
u_\pm  = \frac{1}{2} \left(r_0^2 - 2a^2 \pm r_0 \sqrt{r_0^2-4a^2} \right) \ .
\end{align}
For $4 a^2 = r_0^2$ the horizons merge and the space-time becomes extreme. 

The ergosphere defined by $g_{tt} = 0$ is located at 
\begin{align}
u_{\rm{stat}} = r_0^2 - a^2.
\end{align}
and does not depend on the angle $\vartheta$ for $a=b$.

For test particles or light ($g_{\mu\nu} u^\mu u^\nu = - \delta$,
$\delta = 1$ for massive test particles and $\delta = 0$ for light) 
moving on a geodesic the energy $E$ and angular momenta $\Phi$ and $\Psi$ are conserved.

It is straightforward to see that in the Myers-Perry space--time in 5 dimensions the Hamilton--Jacobi equation 
\begin{equation}
- 2\frac{\partial S}{\partial\tau} = g^{ \mu\nu}\frac{\partial S}{\partial x^\mu}\frac{\partial S}{\partial x^\nu} \ ,
\end{equation}
where $\tau$ is an affine parameter along the geodesic, is separable and yields for each coordinate a corresponding differential equation
\begin{eqnarray}
\left(\frac{du}{d\gamma}\right)^2 & = & U \label{eq-r-theta:1} \\
\left(\frac{d\vartheta }{d\gamma }\right)^2 & = & \Theta \,  \label{eq-r-theta:2} \\ 
\frac{d\varphi}{d\gamma} & =  & \frac{\Phi}{\sin^2\vartheta} - r_0^2\frac{a\mathcal{E}}{\Delta\alpha} - (a^2-b^2)\frac{\Phi}{\alpha}  \, \label{dvarphidgamma} \\
\frac{d\psi}{d\gamma} & =  & \frac{\Psi}{\cos^2\vartheta} - r_0^2\frac{b\mathcal{E}}{\Delta\beta} + (a^2-b^2)\frac{\Psi}{\beta}  \, \label{dpsidgamma} \\
\frac{d t}{d\gamma} & = & \rho^2 E + r_0^2\frac{\mathcal{E}}{\Delta} \, .  \label{dtdgamma}
\end{eqnarray}
Here we used $\mathcal{E} = E\alpha\beta + a\Phi\beta + b\Psi\alpha $ and introduced the Mino time $\gamma$ through $\rho^2 d\gamma = d \tau$ \cite{Mino03}. We also defined
\begin{eqnarray}
\frac{U}{4} & = & \Delta ( (E^2 - \delta)u - K + V) + r_0^2 \frac{\mathcal{E}^2}{\alpha \beta}  \label{R_polynomial} \\
\Theta & = & (E^2-\delta)(a^2\cos^2\vartheta + b^2\sin^2\vartheta) + K - \frac{\Phi^2}{\sin^2\vartheta} - \frac{\Psi^2}{\cos^2\vartheta}   \label{Theta_polynomial} \, , 
\end{eqnarray}
where $V=(a^2-b^2)\left( \frac{\Phi^2}{\alpha} - \frac{\Psi^2}{\beta} \right)$. 
The separation constant $K$ is known as Carter constant.

In the following chapters we consider the case of equal rotation parameters: $a=b$.

\section{Complete classification of geodesics}\label{theta-r-features}

Consider the Hamilton-Jacobi equations~\eqref{eq-r-theta:1}-\eqref{dtdgamma}. The properties of the orbits are given by the polynomial $U$~\eqref{R_polynomial} and the function $\Theta$~\eqref{Theta_polynomial}. The constants of motion (energy, angular momenta and separation constant) as well as the parameters of the metric (mass, rotation parameter) characterize these polynomials and, as a consequence, the types of orbits. In this section we discuss the motion in Myers-Perry in terms of the properties of the underlying polynomial $U$ and the function $\Theta$.

\begin{table*}[ht]
\hfill{}
\begin{tabular}{lcccl}\hline
Type & Region & Roots in $[-a^2, \infty)$ & $u$--region & Orbit \\ \hline\hline
A & (3) & 1  & 
\begin{pspicture}(-2,-0.2)(3,0.2)
\psline[linewidth=0.5pt]{->}(-2,0)(3,0)
\psline[linewidth=0.5pt,doubleline=true](1.0,-0.2)(1.0,0.2)
\psline[linewidth=0.5pt,doubleline=true](-0.5,-0.2)(-0.5,0.2)
\psline[linewidth=1.5pt](-2,-0.2)(-2,0.2)
\psline[linewidth=1.2pt]{*-}(-1.1,0)(3,0)
\end{pspicture} 
 & TEO \\ 
${\rm A}_-$ & & & 
\begin{pspicture}(-2,-0.2)(3,0.2)
\psline[linewidth=0.5pt]{->}(-2,0)(3,0)
\psline[linewidth=0.5pt,doubleline=true](1.0,-0.2)(1.0,0.2)
\psline[linewidth=0.5pt,doubleline=true](-0.5,-0.2)(-0.5,0.2)
\psline[linewidth=1.5pt](-2,-0.2)(-2,0.2)
\psline[linewidth=1.2pt]{*-}(-0.5,0)(3,0)
\end{pspicture} 
 & ${\rm TEO}_-$ \\
  ${\rm A}_{\rm s}$ &   &   & 
\begin{pspicture}(-2,-0.2)(3,0.2)
\psline[linewidth=0.5pt]{->}(-2,0)(3,0)
\psline[linewidth=0.5pt,doubleline=true](1.0,-0.2)(1.0,0.2)
\psline[linewidth=0.5pt,doubleline=true](-0.5,-0.2)(-0.5,0.2)
\psline[linewidth=1.5pt](-2,-0.2)(-2,0.2)
\psline[linewidth=1.2pt]{*-}(-2,0)(3,0)
\end{pspicture} 
 & ${\rm TEO}_s$ \\ \hline
B & $(1)$ & 2  & 
\begin{pspicture}(-2,-0.2)(3,0.2)
\psline[linewidth=0.5pt]{->}(-2,0)(3,0)
\psline[linewidth=0.5pt,doubleline=true](1.0,-0.2)(1.0,0.2)
\psline[linewidth=0.5pt,doubleline=true](-0.5,-0.2)(-0.5,0.2)
\psline[linewidth=1.5pt](-2,-0.2)(-2,0.2)
\psline[linewidth=1.2pt]{*-*}(-1.0,0)(1.5,0)
\end{pspicture} 
& MBO \\  
${\rm B}_+$ & &  & 
\begin{pspicture}(-2,-0.2)(3,0.2)
\psline[linewidth=0.5pt]{->}(-2,0)(3,0)
\psline[linewidth=0.5pt,doubleline=true](1.0,-0.2)(1.0,0.2)
\psline[linewidth=0.5pt,doubleline=true](-0.5,-0.2)(-0.5,0.2)
\psline[linewidth=1.5pt](-2,-0.2)(-2,0.2)
\psline[linewidth=1.2pt]{*-*}(-0.8,0)(1,0)
\end{pspicture} 
& ${\rm MBO}_+$ \\ 
${\rm B}_-$ & &  & 
\begin{pspicture}(-2,-0.2)(3,0.2)
\psline[linewidth=0.5pt]{->}(-2,0)(3,0)
\psline[linewidth=0.5pt,doubleline=true](1.0,-0.2)(1.0,0.2)
\psline[linewidth=0.5pt,doubleline=true](-0.5,-0.2)(-0.5,0.2)
\psline[linewidth=1.5pt](-2,-0.2)(-2,0.2)
\psline[linewidth=1.2pt]{*-*}(-0.5,0)(1.3,0)
\end{pspicture} 
& ${\rm MBO}_-$ \\ 
${\rm B}_\pm$ & &  & 
\begin{pspicture}(-2,-0.2)(3,0.2)
\psline[linewidth=0.5pt]{->}(-2,0)(3,0)
\psline[linewidth=0.5pt,doubleline=true](1.0,-0.2)(1.0,0.2)
\psline[linewidth=0.5pt,doubleline=true](-0.5,-0.2)(-0.5,0.2)
\psline[linewidth=1.5pt](-2,-0.2)(-2,0.2)
\psline[linewidth=1.2pt]{*-*}(-0.5,0)(1.0,0)
\end{pspicture} 
& ${\rm MBO}_\pm$ \\
${\rm B}_{\rm s}$ & &  & 
\begin{pspicture}(-2,-0.2)(3,0.2)
\psline[linewidth=0.5pt]{->}(-2,0)(3,0)
\psline[linewidth=0.5pt,doubleline=true](1.0,-0.2)(1.0,0.2)
\psline[linewidth=0.5pt,doubleline=true](-0.5,-0.2)(-0.5,0.2)
\psline[linewidth=1.5pt](-2,-0.2)(-2,0.2)
\psline[linewidth=1.2pt]{*-*}(-2,0)(1.5,0)
\end{pspicture} 
& ${\rm TBO}$ \\ 
   ${\rm B}_{\rm s+}$ &  &  & 
\begin{pspicture}(-2,-0.2)(3,0.2)
\psline[linewidth=0.5pt]{->}(-2,0)(3,0)
\psline[linewidth=0.5pt,doubleline=true](1.0,-0.2)(1.0,0.2)
\psline[linewidth=0.5pt,doubleline=true](-0.5,-0.2)(-0.5,0.2)
\psline[linewidth=1.2pt]{*-*}(-2.0,0)(1,0)
\psline[linewidth=1.5pt](-2,-0.2)(-2,0.2)
\psline[linewidth=0.5pt]{-}(2.0,0)(3,0)
\end{pspicture} 
 & ${\rm TBO}_{+}$ \\  \hline
C & $(2)$ & 3 & 
\begin{pspicture}(-2,-0.2)(3,0.2)
\psline[linewidth=0.5pt]{->}(-2,0)(3,0)
\psline[linewidth=0.5pt,doubleline=true](1.0,-0.2)(1.0,0.2)
\psline[linewidth=0.5pt,doubleline=true](-0.5,-0.2)(-0.5,0.2)
\psline[linewidth=1.2pt]{*-*}(-1.0,0)(1.5,0)
\psline[linewidth=1.5pt](-2,-0.2)(-2,0.2)
\psline[linewidth=1.2pt]{*-}(2.0,0)(3,0)
\end{pspicture} 
 & MBO, EO \\ 
 ${\rm C}_-$ & & & 
\begin{pspicture}(-2,-0.2)(3,0.2)
\psline[linewidth=0.5pt]{->}(-2,0)(3,0)
\psline[linewidth=0.5pt,doubleline=true](1.0,-0.2)(1.0,0.2)
\psline[linewidth=0.5pt,doubleline=true](-0.5,-0.2)(-0.5,0.2)
\psline[linewidth=1.2pt]{*-*}(-0.5,0)(1.5,0)
\psline[linewidth=1.5pt](-2,-0.2)(-2,0.2)
\psline[linewidth=1.2pt]{*-}(2.0,0)(3,0)
\end{pspicture} 
 &${\rm MBO}_-$, EO \\ 
 ${\rm C}_+$ &  &  & 
\begin{pspicture}(-2,-0.2)(3,0.2)
\psline[linewidth=0.5pt]{->}(-2,0)(3,0)
\psline[linewidth=0.5pt,doubleline=true](1.0,-0.2)(1.0,0.2)
\psline[linewidth=0.5pt,doubleline=true](-0.5,-0.2)(-0.5,0.2)
\psline[linewidth=1.2pt]{*-*}(-1.0,0)(1,0)
\psline[linewidth=1.5pt](-2,-0.2)(-2,0.2)
\psline[linewidth=1.2pt]{*-}(2.0,0)(3,0)
\end{pspicture} 
 & ${\rm MBO}_+$, EO \\ 
  ${\rm C}_{\rm s}$ &  &  & 
\begin{pspicture}(-2,-0.2)(3,0.2)
\psline[linewidth=0.5pt]{->}(-2,0)(3,0)
\psline[linewidth=0.5pt,doubleline=true](1.0,-0.2)(1.0,0.2)
\psline[linewidth=0.5pt,doubleline=true](-0.5,-0.2)(-0.5,0.2)
\psline[linewidth=1.2pt]{*-*}(-2.0,0)(1.5,0)
\psline[linewidth=1.5pt](-2,-0.2)(-2,0.2)
\psline[linewidth=1.2pt]{*-}(2.0,0)(3,0)
\end{pspicture} 
 & ${\rm TBO}$, EO \\ 
   ${\rm C}_{\rm s+}$ &  &  & 
\begin{pspicture}(-2,-0.2)(3,0.2)
\psline[linewidth=0.5pt]{->}(-2,0)(3,0)
\psline[linewidth=0.5pt,doubleline=true](1.0,-0.2)(1.0,0.2)
\psline[linewidth=0.5pt,doubleline=true](-0.5,-0.2)(-0.5,0.2)
\psline[linewidth=1.2pt]{*-*}(-2.0,0)(1,0)
\psline[linewidth=1.5pt](-2,-0.2)(-2,0.2)
\psline[linewidth=1.2pt]{*-}(2.0,0)(3,0)
\end{pspicture} 
 & ${\rm TBO}_{+}$, EO \\  \hline
${\rm D}$ & $(4)$ & 3 &  
\begin{pspicture}(-2,-0.2)(3,0.2)
\psline[linewidth=0.5pt]{->}(-2,0)(3,0)
\psline[linewidth=0.5pt,doubleline=true](1.0,-0.2)(1.0,0.2)
\psline[linewidth=0.5pt,doubleline=true](-0.5,-0.2)(-0.5,0.2)
\psline[linewidth=1.2pt]{*-*}(-1.6,0)(-1.1,0)
\psline[linewidth=1.5pt](-2,-0.2)(-2,0.2)
\psline[linewidth=1.2pt]{*-}(-0.8,0)(3,0)
\end{pspicture} 
 & BO, TEO \\
${\rm D}_-$ &  &  & 
\begin{pspicture}(-2,-0.2)(3,0.2)
\psline[linewidth=0.5pt]{->}(-2,0)(3,0)
\psline[linewidth=0.5pt,doubleline=true](1.0,-0.2)(1.0,0.2)
\psline[linewidth=0.5pt,doubleline=true](-0.5,-0.2)(-0.5,0.2)
\psline[linewidth=1.2pt]{*-*}(-1.6,0)(-1.1,0)
\psline[linewidth=1.5pt](-2,-0.2)(-2,0.2)
\psline[linewidth=1.2pt]{*-}(-0.5,0)(3,0)
\end{pspicture} 
 & BO, ${\rm TEO}_-$\\
 ${\rm D}_{\rm s}$ &  &  & 
\begin{pspicture}(-2,-0.2)(3,0.2)
\psline[linewidth=0.5pt]{->}(-2,0)(3,0)
\psline[linewidth=0.5pt,doubleline=true](1.0,-0.2)(1.0,0.2)
\psline[linewidth=0.5pt,doubleline=true](-0.5,-0.2)(-0.5,0.2)
\psline[linewidth=1.2pt]{*-*}(-2,0)(-1.5,0)
\psline[linewidth=1.5pt](-2,-0.2)(-2,0.2)
\psline[linewidth=1.2pt]{*-}(-0.8,0)(3,0)
\end{pspicture} 
 & ${\rm TBO}$, TEO \\ \hline \hline   
 \end{tabular}
 \hfill{}
\caption{Types of orbits in the Myers-Perry space-time for the rotation parameter $a=b$. The horizontal bold lines represent the orbits and the bold points - the turning points. The thin vertical double dashed line represents the horizons and the thick vertical dashed line -- the singularity at $u=-a^2$. In special cases the turning points lie at the horizons ($\pm$), or lead into the singularity (s). The regions (1)-(4) are related to the $E$-$\Phi$ diagrams for the $u$-motion in Fig.~\ref{fig:xroots}.
\label{tab:orbits}}
\end{table*}

\subsection{The $\vartheta$--motion}\label{subsec:theta-pot}

We rewrite the function $\Theta$~\eqref{Theta_polynomial} in the following form:
\begin{eqnarray}
 \Theta &=& (E^2-\delta)a^2 + {K} - ({\Phi}-\Psi)^2 - \frac{({\Phi}\cos^2\vartheta + \Psi\sin^2\vartheta)^2}{\cos^2\vartheta\sin^2\vartheta} \nonumber \\ 
& =& (E^2-\delta)a^2 + {K} - ({\Phi}+\Psi)^2 - \frac{({\Phi}\cos^2\vartheta - \Psi\sin^2\vartheta)^2}{\cos^2\vartheta\sin^2\vartheta}  \ . \qquad \label{Theta_polynomial_1}
\end{eqnarray}
In order to obtain from Eq.~\eqref{eq-r-theta:2} real values 
of the coordinate $\vartheta$ we have to require $\Theta\geq 0$. 
This implies
\begin{equation}
\begin{array}{ll} c_1: = ({E}^2-\delta)a^2 + {K} - ({\Phi}-\Psi)^2 & \geq 0 \\
                  c_2: = ({E}^2-\delta)a^2 + {K} - ({\Phi}+\Psi)^2 &  \geq 0 \ . \end{array} \  \label{theta-cond-lamu} 
\end{equation}
With the new variable $\xi := \cos^2\vartheta$, Eq.~(\ref{eq-r-theta:2}) 
turns into the equation 
\begin{equation}
\left(\frac{d\xi}{d\gamma}\right)^2 = \Theta_\xi \quad \text{with} \quad \Theta_\xi := a_\xi \xi^2 + b_\xi \xi + c_\xi \, , \label{xieom}
\end{equation}
with a simple polynomial of second order on the right hand side, where 
\begin{eqnarray}
 \frac{a_\xi}{4} &=& - (c_1 + ( {\Phi} - \Psi )^2) = - (c_2 + ( {\Phi} + \Psi )^2) \ , \\ 
 \frac{b_\xi}{4} &=& K +a^2({E}^2-\delta)+ \Psi^2 - {\Phi}^2  \ , \\ 
 \frac{c_\xi}{4} &=& - \Psi^2 \ .
\end{eqnarray}
Since $ c_1 \geq 0$ we have $a_\xi < 0$. This means that $\Theta_\xi$ can be positive if and only if there are real zeros of $\Theta_\xi$. The polynomial $\Theta_\xi$ plays the role of an effective potential for the $\vartheta$--motion. The real and positive zeros of $\Theta_\xi$ define the angles of two cones which confine the motion of the test particles. (A similar feature appears in Taub-NUT, Kerr-(de Sitter) and Reissner-Nordstr\"om (for charged particles) space--times~\cite{KKHL10,HLKK10,GK10}.) 

The discriminant $D = b_\xi^2 - 4 a_\xi c_\xi$ of the polynomial $\Theta_\xi$ 
can be written as $D = 16 c_1 c_2$. The existence of real zeros of $\Theta_\xi$ requires $D \geq 0$.
And with the conditions~\eqref{theta-cond-lamu} this is fulfilled. The inequalities~\eqref{theta-cond-lamu} impose limitations on the parameters $E$, ${\Phi}$, $\Psi$ and ${K}$ for some given ${a}$. 

The polynomial $\Theta_\xi(\xi)$ in~\eqref{xieom} has roots $ \xi_{1,2} = - \frac{1}{2 a_\xi} \left(b_\xi \pm \sqrt{D} \right)$, and 
describes a parabola with the maximum at $\left(- \frac{b_\xi}{2 a_\xi}, -\frac{ D}{4a_\xi} \right)$. In special cases the roots $\xi_{1,2}$ are:
\begin{enumerate}
\item  $c_1=0$  $\Rightarrow$ $\xi_{1,2} = \frac{\Psi}{\Psi - \Phi}$. For $\Phi=-\Psi$ the orbit lies at $\vartheta=\frac{\pi}{4}$. 
\item $c_2=0$  $\Rightarrow$ $\xi_{1,2} = \frac{\Psi}{\Psi + \Phi}$. For $\Phi=\Psi$ the orbit lies at $\vartheta=\frac{\pi}{4}$. 
\item  $c_1=c_2=0$  $\Rightarrow$ 2 cases are possible: $\xi_{1,2} = 1$ for $\Phi=0$ and $\xi_{1,2} = 0$ for $\Psi=0$. Thus, the motion is in the $\vartheta=0$ plane or $\vartheta=\frac{\pi}{2}$ plane.
\end{enumerate}

\begin{figure*}[th!]
\begin{center}
\subfigure[][$a=0, K=3, \Psi=0.5$.]{\label{theta0a}\includegraphics[width=4.5cm]{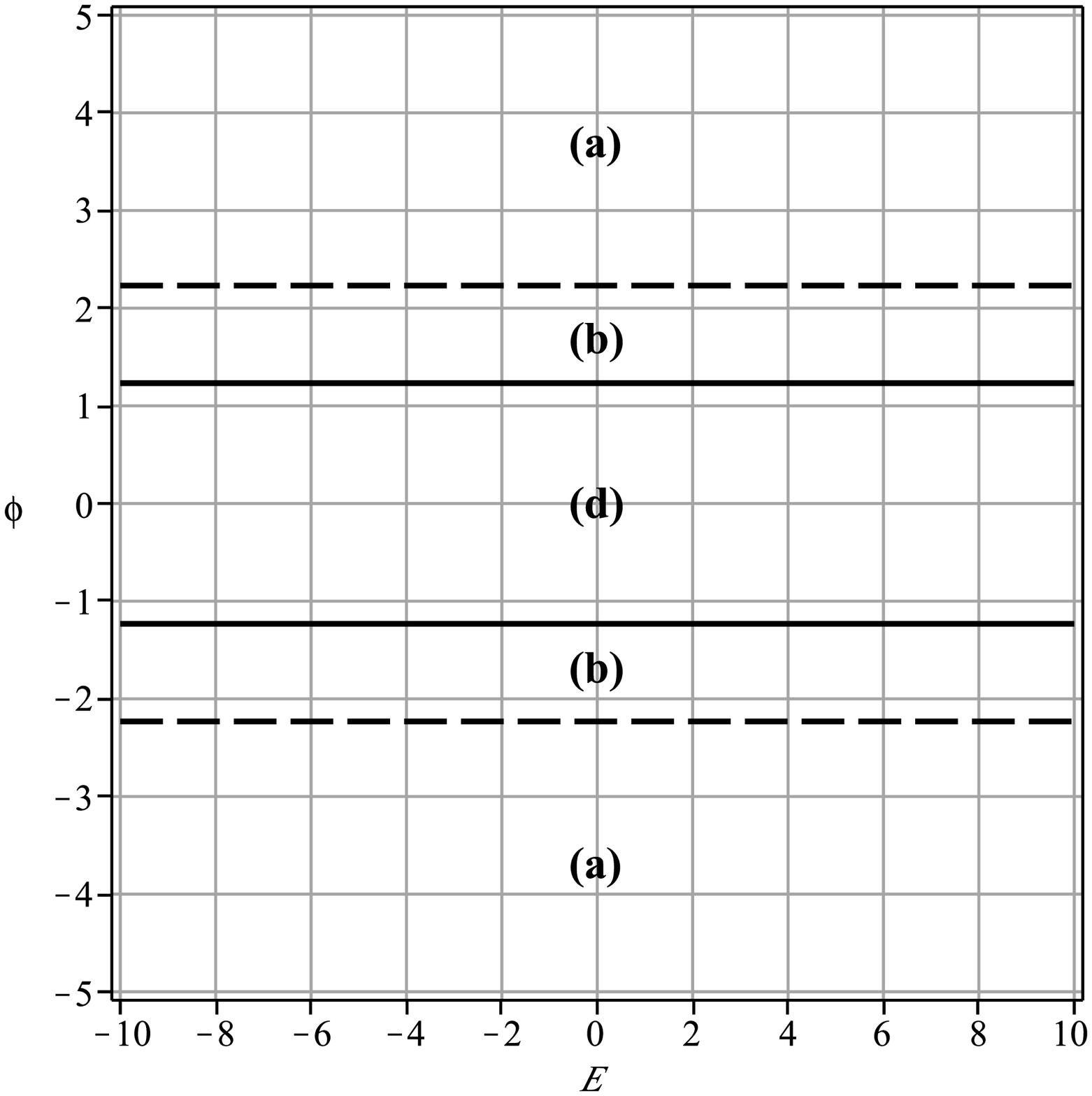}}
\subfigure[][$a=0.2, K=3, \Psi=0.5$.]{\label{theta0b}\includegraphics[width=4.5cm]{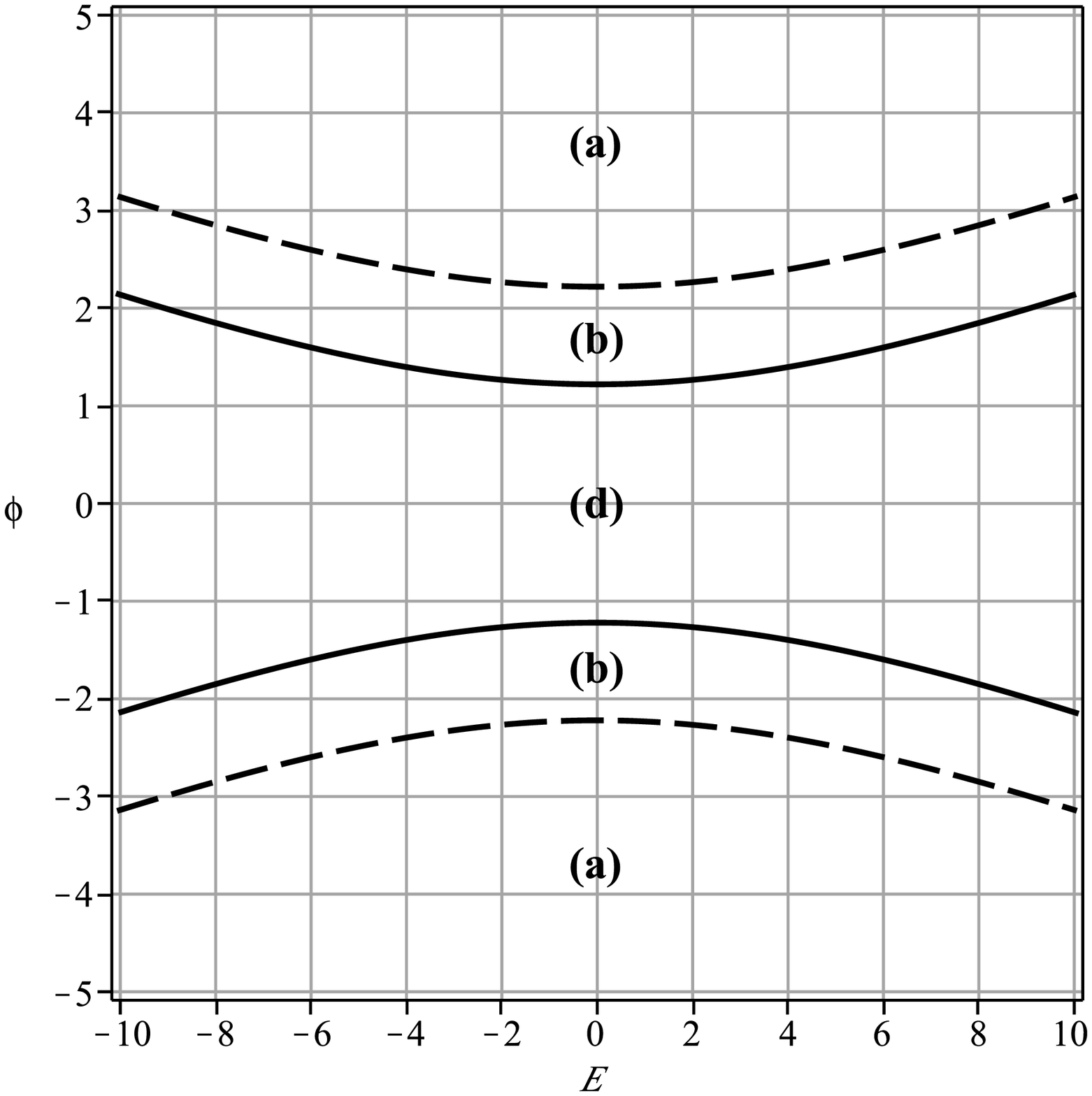}}
\subfigure[][$a=0.4, K=3, \Psi=0.5$.]{\label{theta0c}\includegraphics[width=4.5cm]{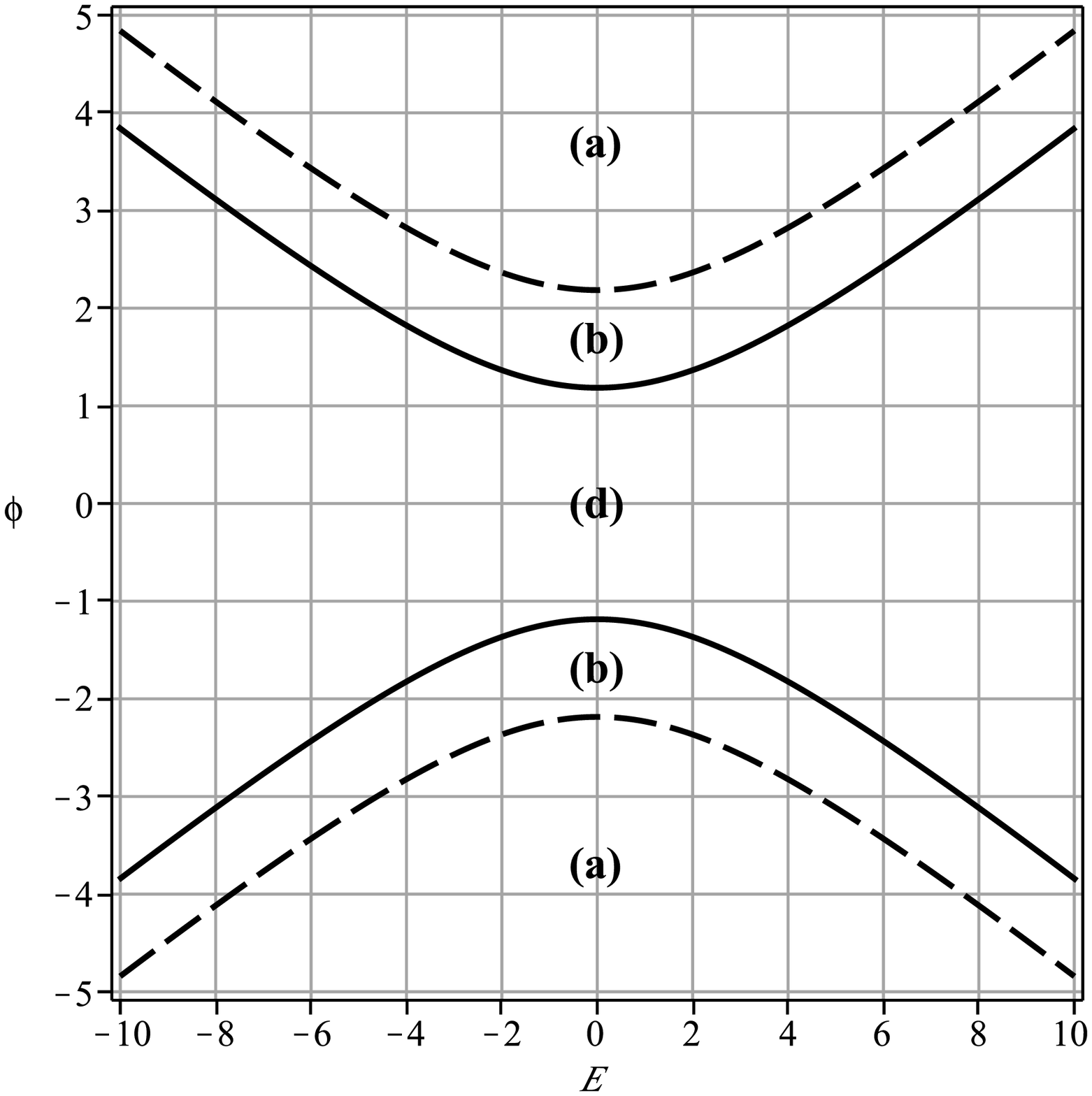}} \\
\subfigure[][$a=0.4, K=3, \Psi=0$.]{\label{theta0d}\includegraphics[width=4.5cm]{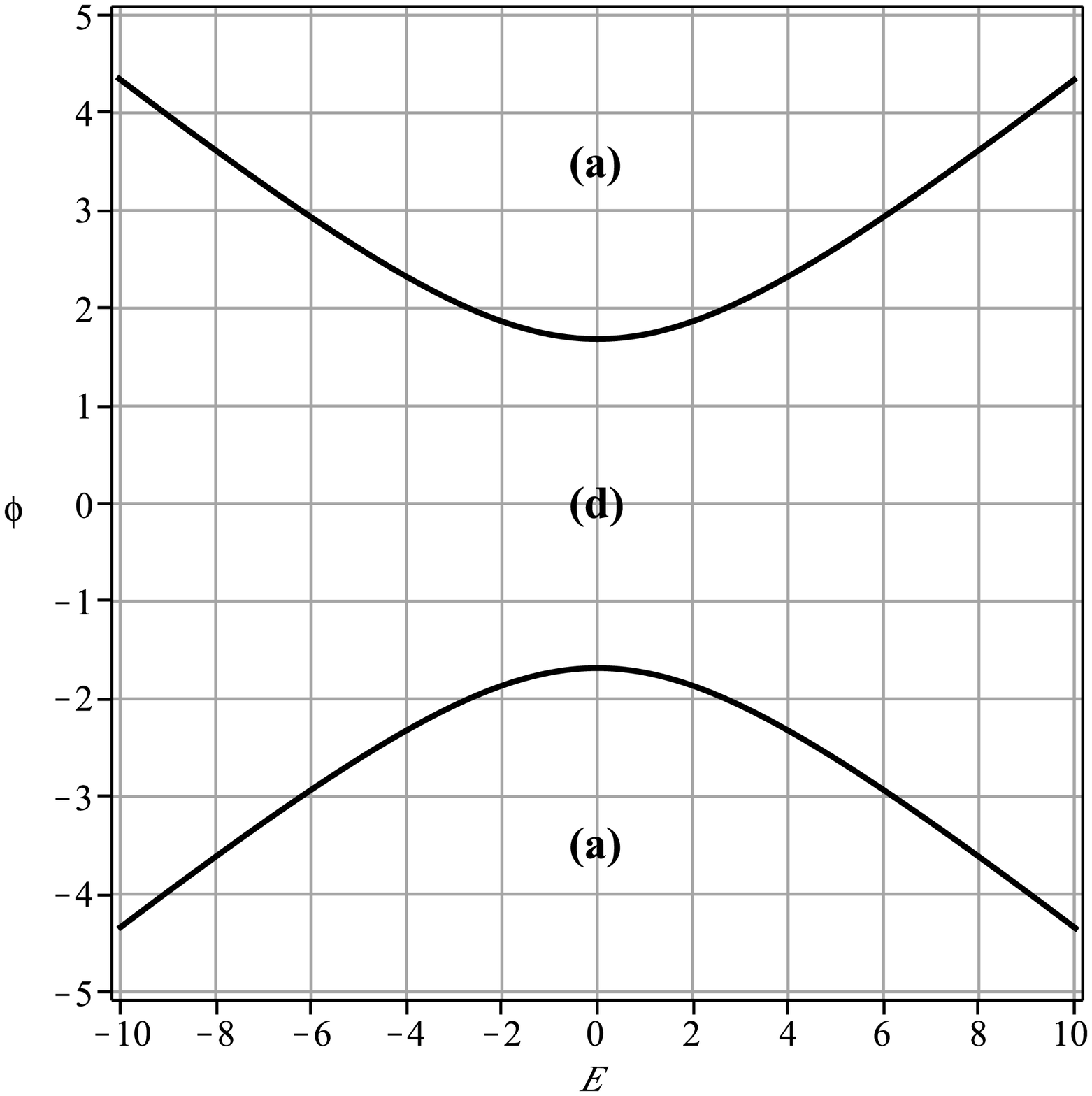}}
\subfigure[][$a=0.4, K=3, \Psi=1.5$.]{\label{theta0e}\includegraphics[width=4.5cm]{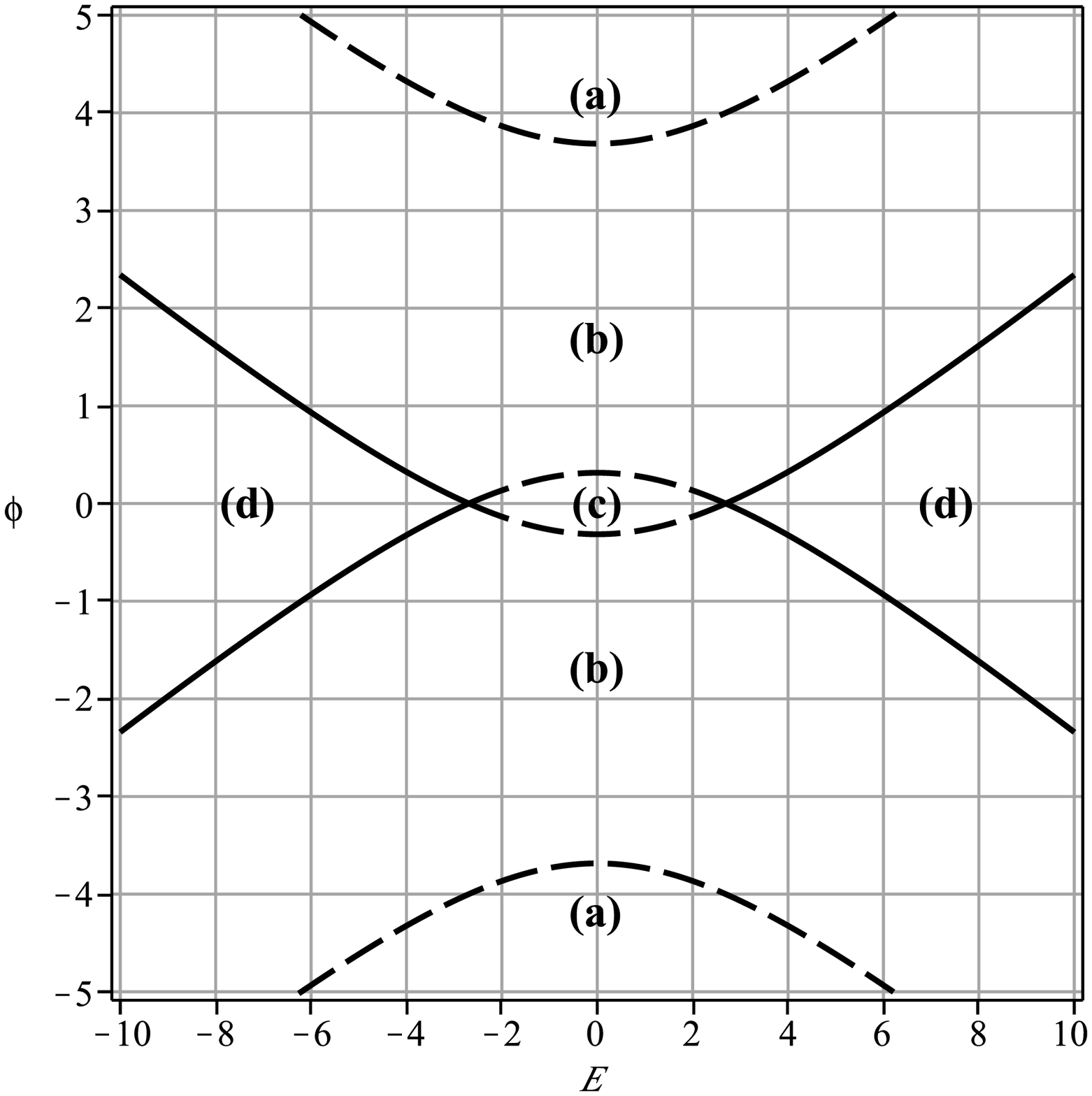}}
\subfigure[][$a=0.4, K=10, \Psi=0.5$.]{\label{theta0f}\includegraphics[width=4.5cm]{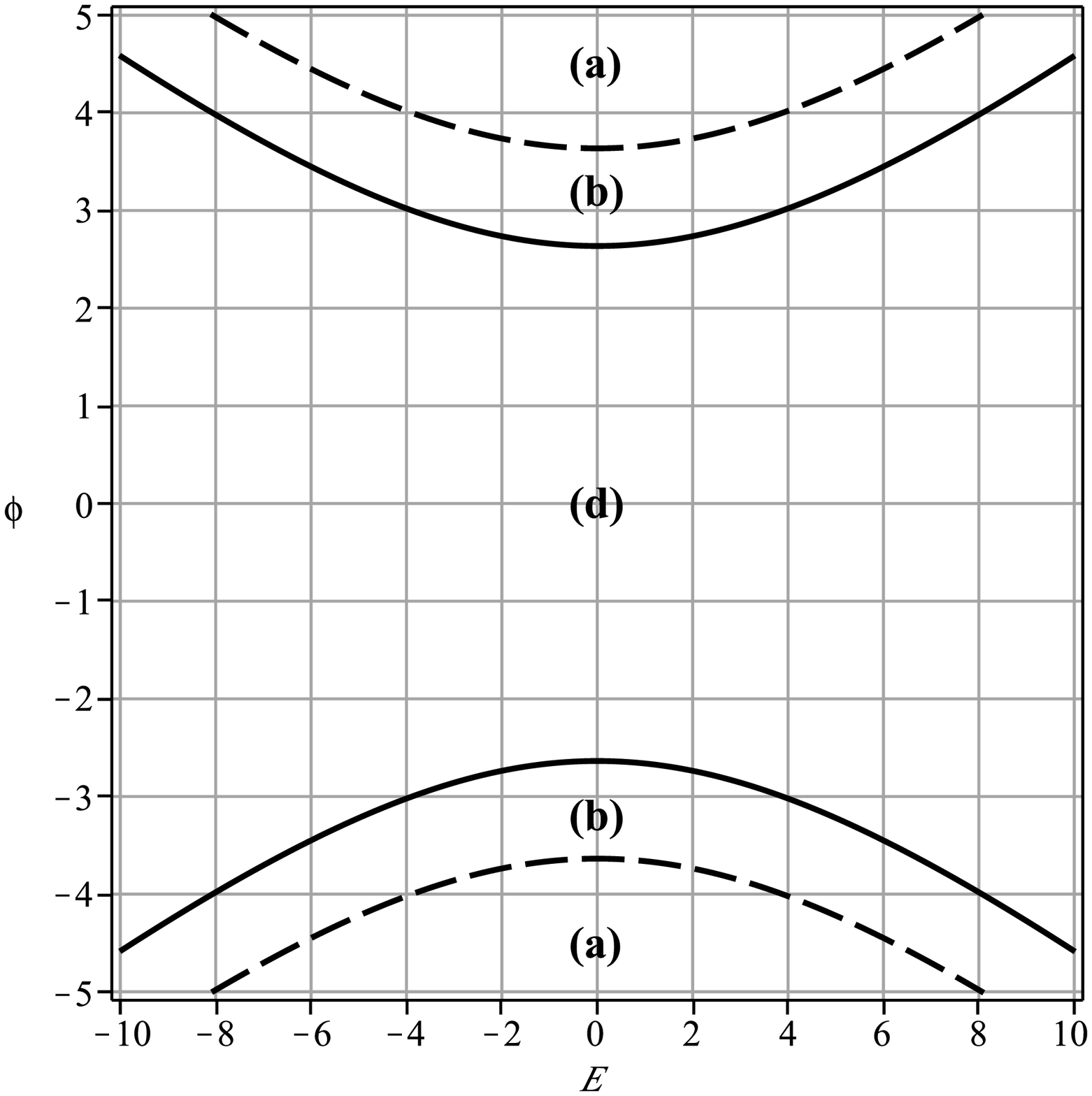}}
\end{center}
\caption{The $E$-$\Phi$ diagram for the $\vartheta$-motion for different values of $a$, $K$, $\Phi$ and $\Psi$. In the region (b) the discriminant is negative $D<0$. In the regions (a), (d) and in some cases in the region (c) the discriminant is positive $D_\xi>0$, but only in the region (d) the conditions~\eqref{theta-cond-lamu} are fulfilled. On the lines the discriminant vanishes. Moreover, on the dashed lines one of the quantities $c_{1,2}$ is negative and the motion is not allowed.  \label{fig:thetaroots}}
\end{figure*}

Fig.~\ref{fig:thetaroots} shows the $\Phi$-$E$ dependence on the parameters $a$, $K$, $\Psi$ from the function~\eqref{Theta_polynomial_1}. Only in the region $\rm{(d)}$ $c_1$, $c_2$ and $D$ are positive, in all other regions either $D$ is negative or one of the conditions~\eqref{theta-cond-lamu} or both is/are not fulfilled. On the lines the discriminant $D$ vanishes and the motion is characterized by a constant angle $\vartheta$. For small values of the rotation parameter $a$ the curves are less bent and for $a=0$ the curves become straight lines. For growing values of the separation constant $K$ the distance between the regions $\rm{(b)}$ grows. By varying $\Psi$ we change the size of the region $\rm{(d)}$.

\subsection{The $u$--motion}\label{subsec:r-motion}

The right hand side of the differential equation~\eqref{eq-r-theta:1} has the form $U=\sum^3_{i=0}{b_i u^i}$ with the coefficients $b_i$ for $a=b$:

\begin{align}
\label{eq:33}
\begin{aligned}
b_3 &= 4 \left(E^2 - \delta \right)\\[8pt]
b_2 &=8 a^2 \left(E^2 - \delta \right) + 4 \left( \delta r_0^2  - K \right) \\[8pt]
b_1 &= 4 a^4 \left(E^2 - \delta \right) + 4 K (r_0^2 - 2 a^2) + 8 r_0^2 (a^2 E^2 + a E (\Phi+\Psi))
 \\[8pt]
b_0 &= 4 a^2 \left( r_0^2  \left(  a E + \Psi + \Psi \right)^2 - K a^2  \right) .
\end{aligned}
\end{align}

The list of the possible orbits reads (For an explanation of the names of the orbits see~\cite{Hackmannetal08}):

\begin{enumerate}
     \item \textit{Escape Orbits} (EO) in the region $[u_1, \infty)$ with $u_1 > u_+$. The EO does not cross the horizons.\vspace{4pt}
     \item \textit{Two-world escape orbits} (TEO) in the region $[u_1, \infty)$ with $|u_1| \leq u_-$ 
      \begin{enumerate}
     \item \textit{Terminating escape orbits} (TEO$_{\rm s}$) in the region $[-a^2, \infty)$. The test particle comes from infinity and disappears into the singularity ($u=-a^2$). \vspace{4pt}
 \end{enumerate}
     \item \textit{Periodic bound orbits} (BO) in the region $u \in [u_1, u_2]$ with $u_1 > u_2$ and $u_1, u_2 < u_-$ (planetary orbits).
       \begin{enumerate}
     \item \textit{Terminating bound orbit} (TBO) in the region $u \in [-a^2, u_1]$ with $u_1 < u_-$. The turning point of a bound orbit coincides with the singularity for some energy value.\vspace{4pt}
 \end{enumerate}
      \item \textit{Many-world periodic bound orbits} (MBO) in the region $u \in [u_1, u_2]$ with $u_1 \le u_-$ and $u_2 \ge u_+$.
        \begin{enumerate}
      \item Also for the MBO it is possible that the turning point concides with the singularity and the orbit turns to a TBO in the region $u \in [-a^2, u_1]$, with $u_1 \ge u_+$.\\
 \end{enumerate}
  \end{enumerate}

\begin{figure*}[th!]
\begin{center}
\subfigure[][$r_0=1, a=0.2, K=3, \Psi=0.5$. Definition of the regions (1), (2) and (3) (see Table~\ref{tab:orbits})]{\label{x0a}\includegraphics[width=4.5cm]{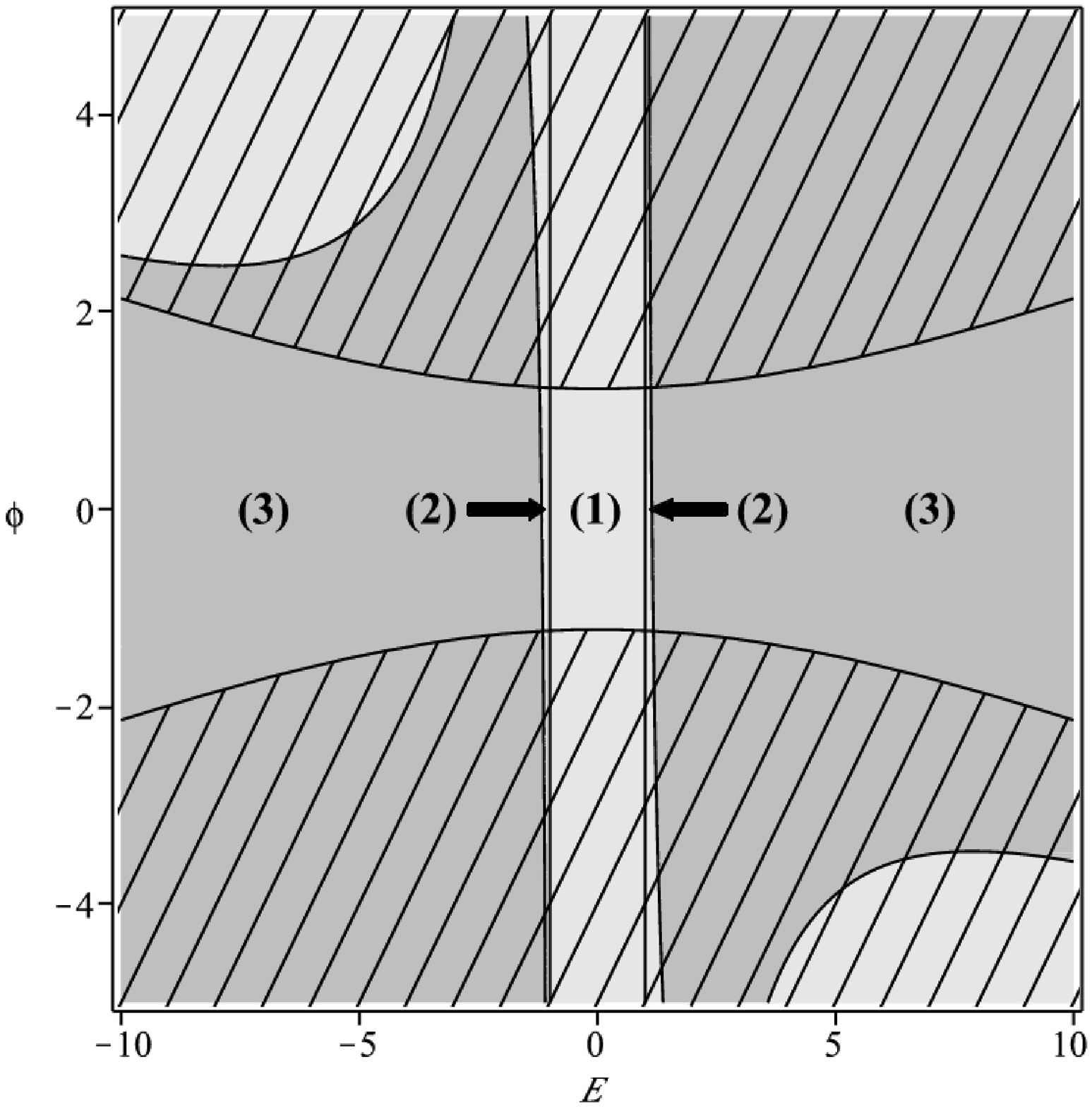}}
\subfigure[][$r_0=1, a=0.4, K=3, \Psi=0.5$.]{\label{x0b}\includegraphics[width=4.5cm]{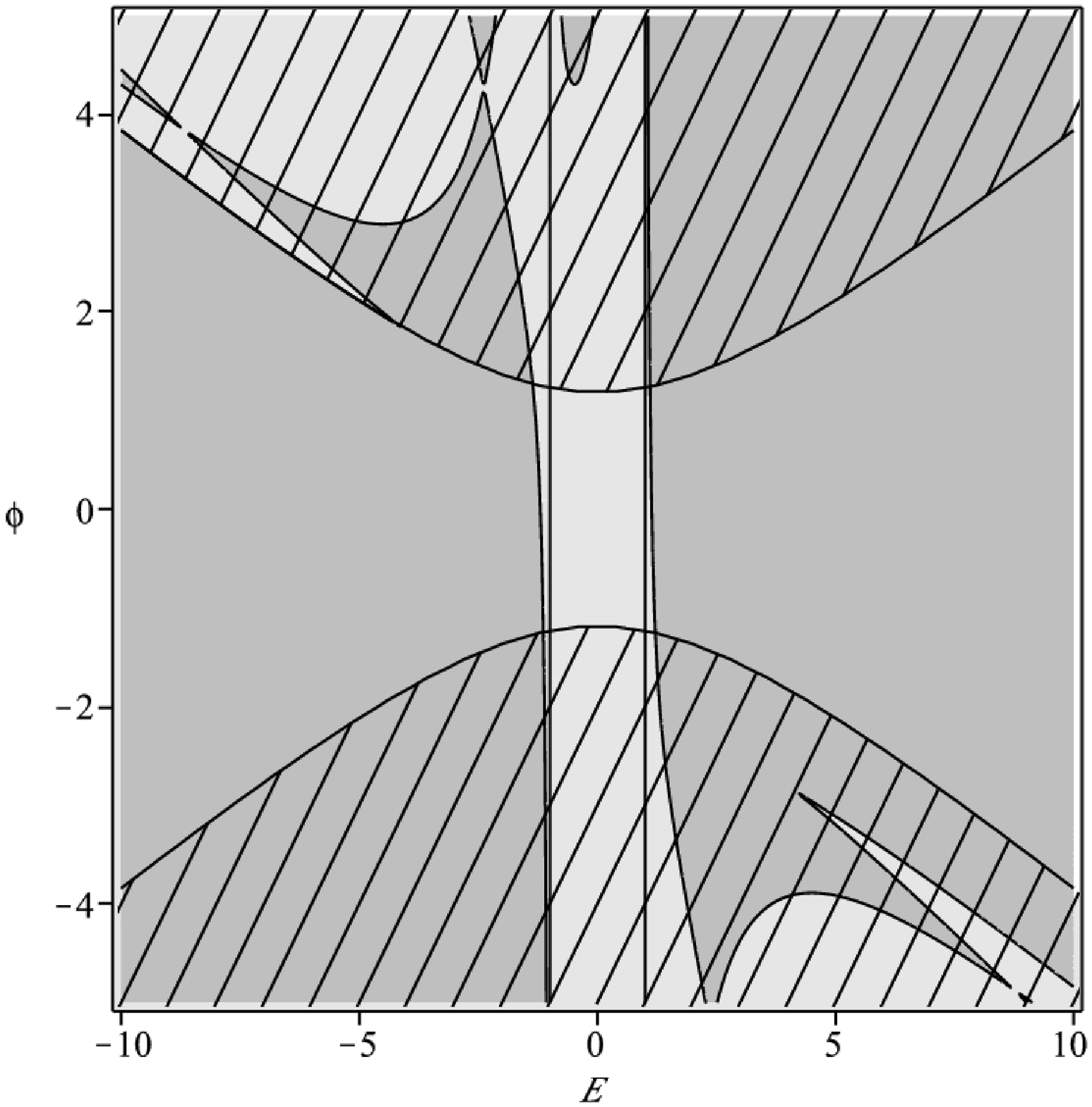}}
\subfigure[][$r_0=1, a=0.4, K=5, \Psi=-0.5$.]{\label{x0c}\includegraphics[width=4.5cm]{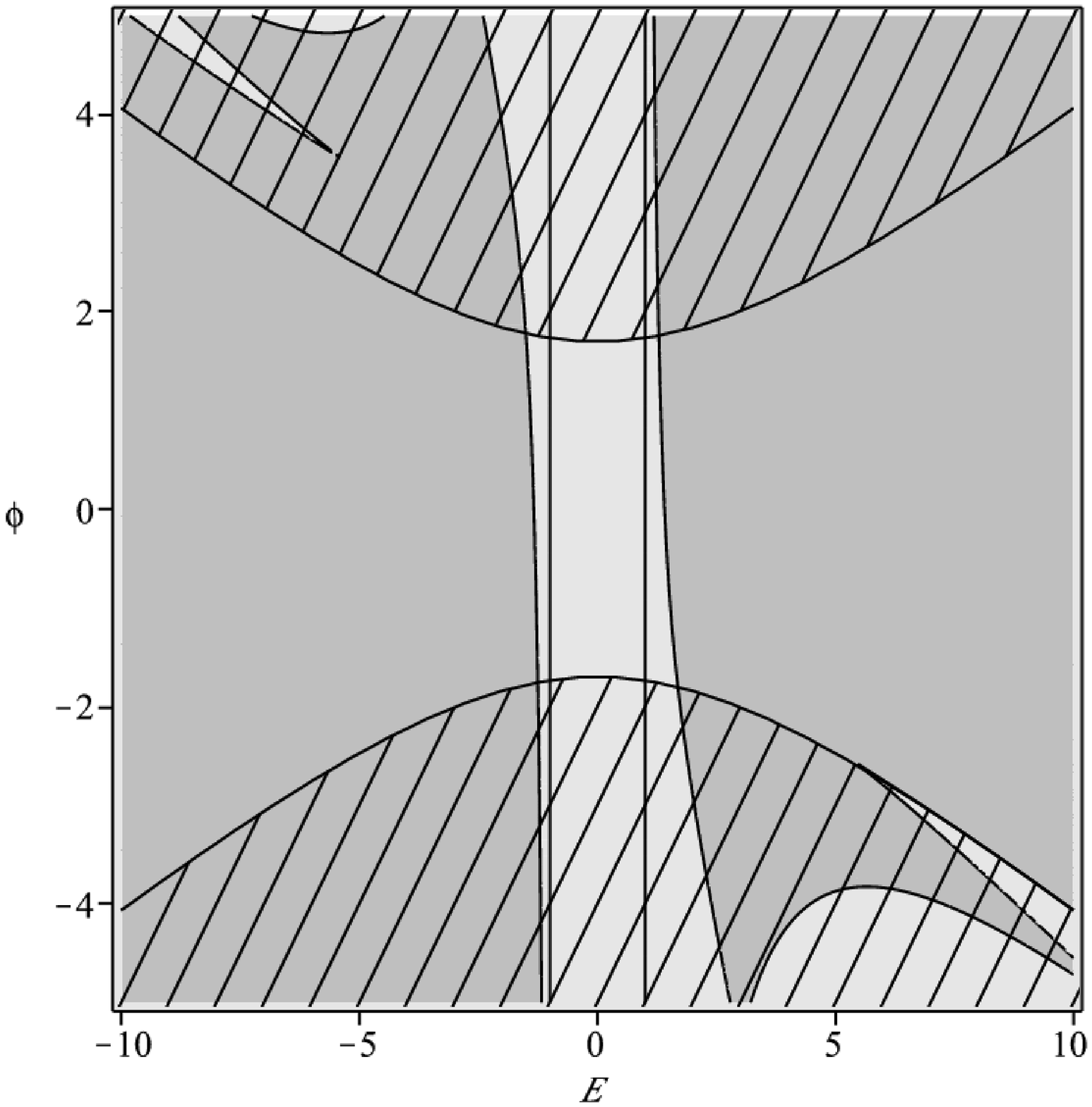}} \\
\subfigure[][$r_0=1, a=0.4, K=3, \Psi=-1.5$.]{\label{x0d}\includegraphics[width=4.5cm]{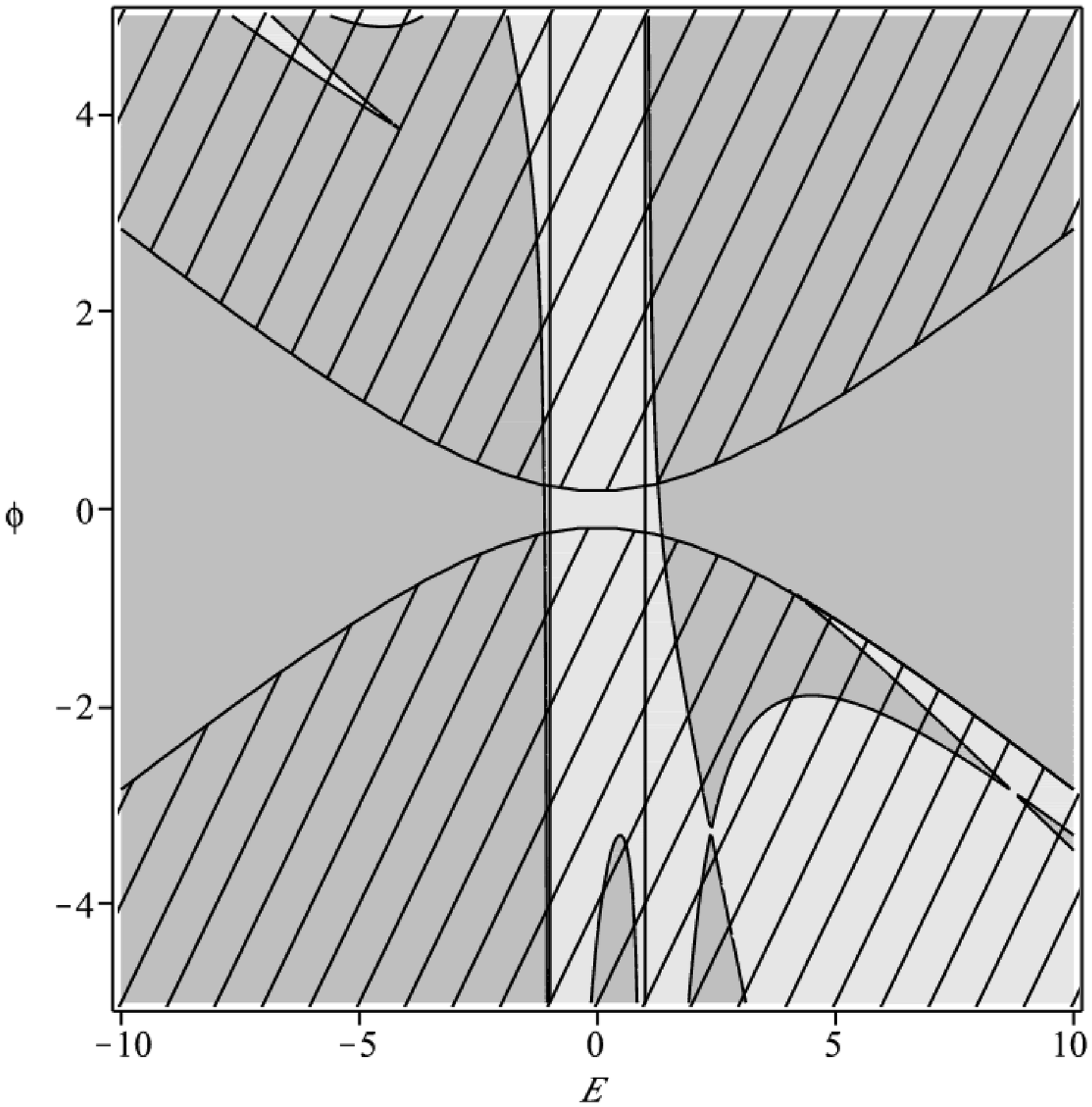}}
\subfigure[][$r_0=1, a=0.4, K=10, \Psi=-4$.]{\label{x0e}\includegraphics[width=4.5cm]{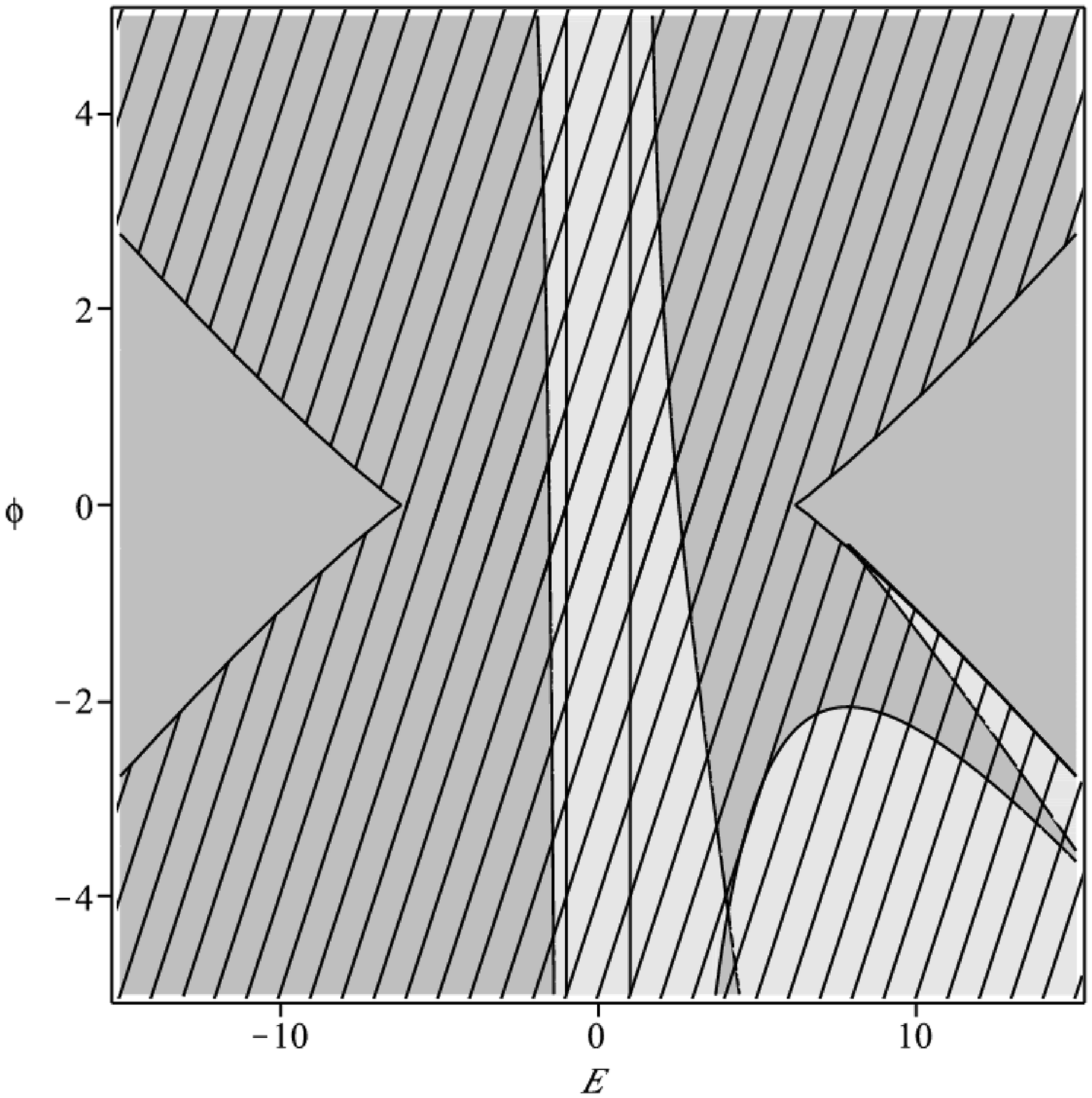}}
\subfigure[][$r_0=1, a=0.4, K=10, \Psi=-4$. Detailed representation of~\subref{x0e}, region (4)]{\label{x0f}\includegraphics[width=4.5cm]{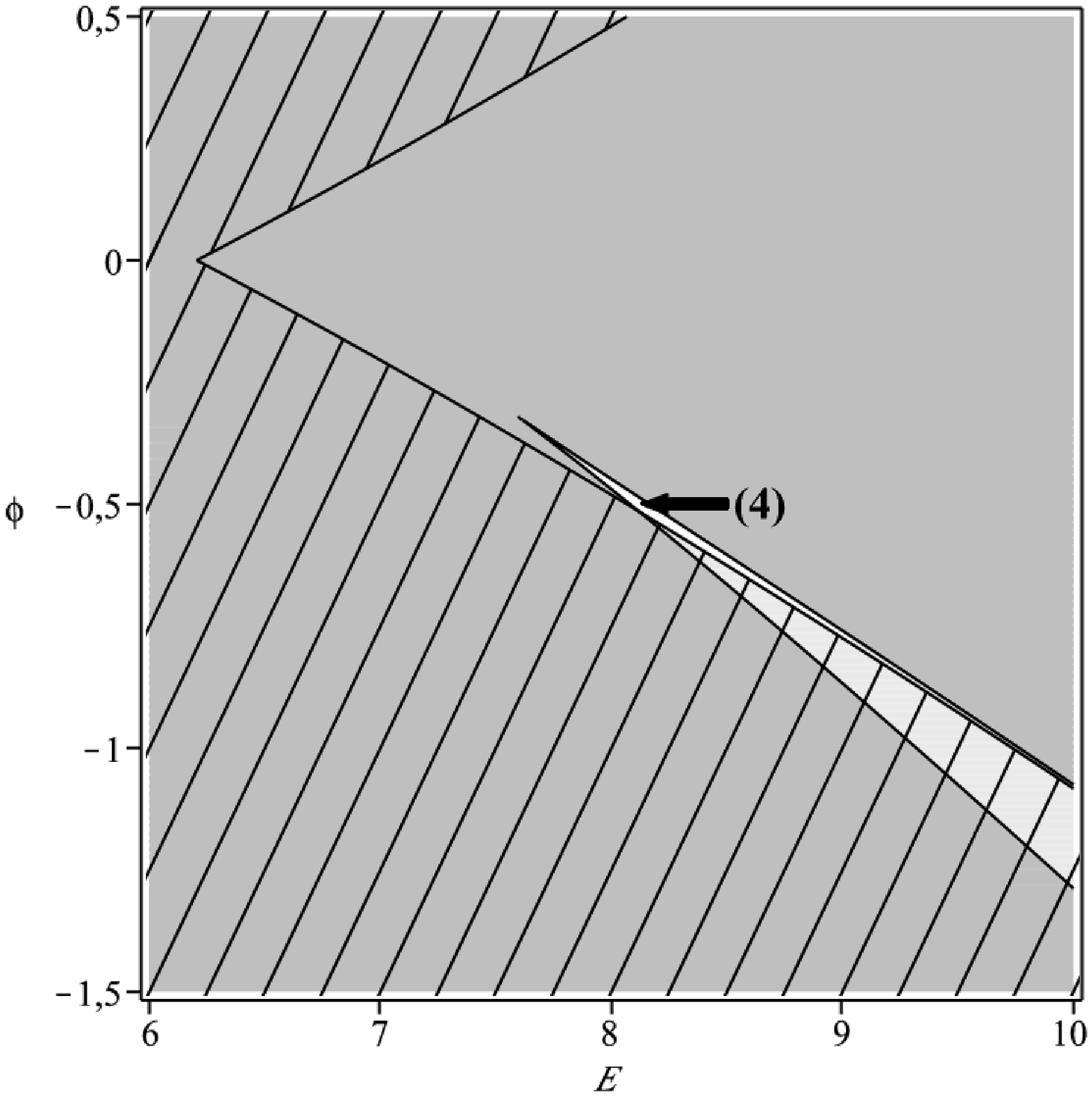}}
\end{center}
\caption{The $E$-$\Phi$ diagrams for the $u$-motion for different values of the parameters $a$, $K$, $\Phi$ and $\Psi$. The light grey region corresponds to three real zeros of the polynomial $U$ and the dark grey denotes the regions with 2 real zeros. When plotting an orbit one has to check in addition which of these real zeros lies in $[-a^2, \infty)$. The dashed region is forbidden by the inequalities~\eqref{theta-cond-lamu}.\label{fig:xroots}}
\end{figure*}

Analysing the polynomial $U$ in the equation~\eqref{eq-r-theta:1} we construct the $E$-$\Phi$ diagrams in Fig.~\ref{fig:xroots} for the $u$-motion. The diagrams illustrate how the change of the values of the parameters $a$, $K$, $\Phi$ and $\Psi$ influences the number of real zeros of the polynomial $U$. From Table~\ref{tab:orbits} we read the possible types of orbits in these regions. In the light grey region are at most 3 real zeros possible. If all three roots belong to the interval $[-a^2, \infty)$, then according to the Table~\ref{tab:orbits} we have regions (2) or region (4) depending on the values of the parameters. It can happen that only 2 of 3 zeros lie in the interval $[-a^2, \infty)$. This corresponds to the region (1). There is only 1 real zero in the region (3) painted in dark grey. The dashed area is forbidden by the inequalities~\eqref{theta-cond-lamu}.

\begin{enumerate}[\text{Region} (1):]
     \item Two real zeros in $[-a^2, \infty)$. According to Table~\ref{tab:orbits} possible orbits are MBO. Special cases are B$_\pm$ when the turning points lie at the horizons, B$_-$ and B$_+$ when one turning point lies at the inner or outer horizon correspondingly.
     A very specific orbit has the left turning point merged with the singularity (B$_{\rm s}$).
     \item Three real zeros in $[-a^2, \infty)$. Possible orbits are MBO und EO with special cases C$_+$, C$_-$, C$_{\rm s}$ and C$_{\rm s+}$.
     \item One real zero in $[-a^2, \infty)$. Possible orbits are TEO for test particles coming from $u= + \infty$. The turning point may lie at the inner horizon (A$_-$), or merge with the singularity (A$_{\rm s}$).
     \item Three real zeros in $[-a^2, \infty)$. Possible orbits are BO and TEO. BO lies behind the inner horizon. The turning point of the TEO may lie at the inner horizon (D$_-$), or the left turning point of the BO may merge with the singularity (D$_{\rm s}$).
  \end{enumerate}

Another useful method to characterise the motion of a test particle is to plot the effective potential $V_{\text{eff}}$. We define it in the following way 
 \begin{align}
 \left( \frac{du}{d\gamma} \right)^2 = 4 \left(\Delta u + \alpha^2 r_0^2\right) \left(E - V_{\text{eff}}^+ \right) \left(E - V_{\text{eff}}^- \right),
 \end{align}

where

\begin{equation}
V_{\text{eff}}^{\pm} = - \frac{1}{\Delta u + \alpha^2 r_0^2} \Bigl(\alpha a r_0^2 \left(\Phi + \Psi \right) \pm \sqrt{\Delta \left(K + \delta u \right) \left(\Delta u + \alpha^2 r_0^2 \right) -  \Delta a^2 r_0^2 u \left(\Phi + \Psi \right)^2} \Bigr) \ .
\end{equation}

\begin{figure*}[th!]
\begin{center}
\subfigure[][$r_0=1, a=0.4, K=5, \Phi=-0.5, \Psi=0.5$. Since $\Phi + \Psi = 0$ the plot is symmetric wrt the u-axis]{\label{pot1}\includegraphics[width=4.5cm]{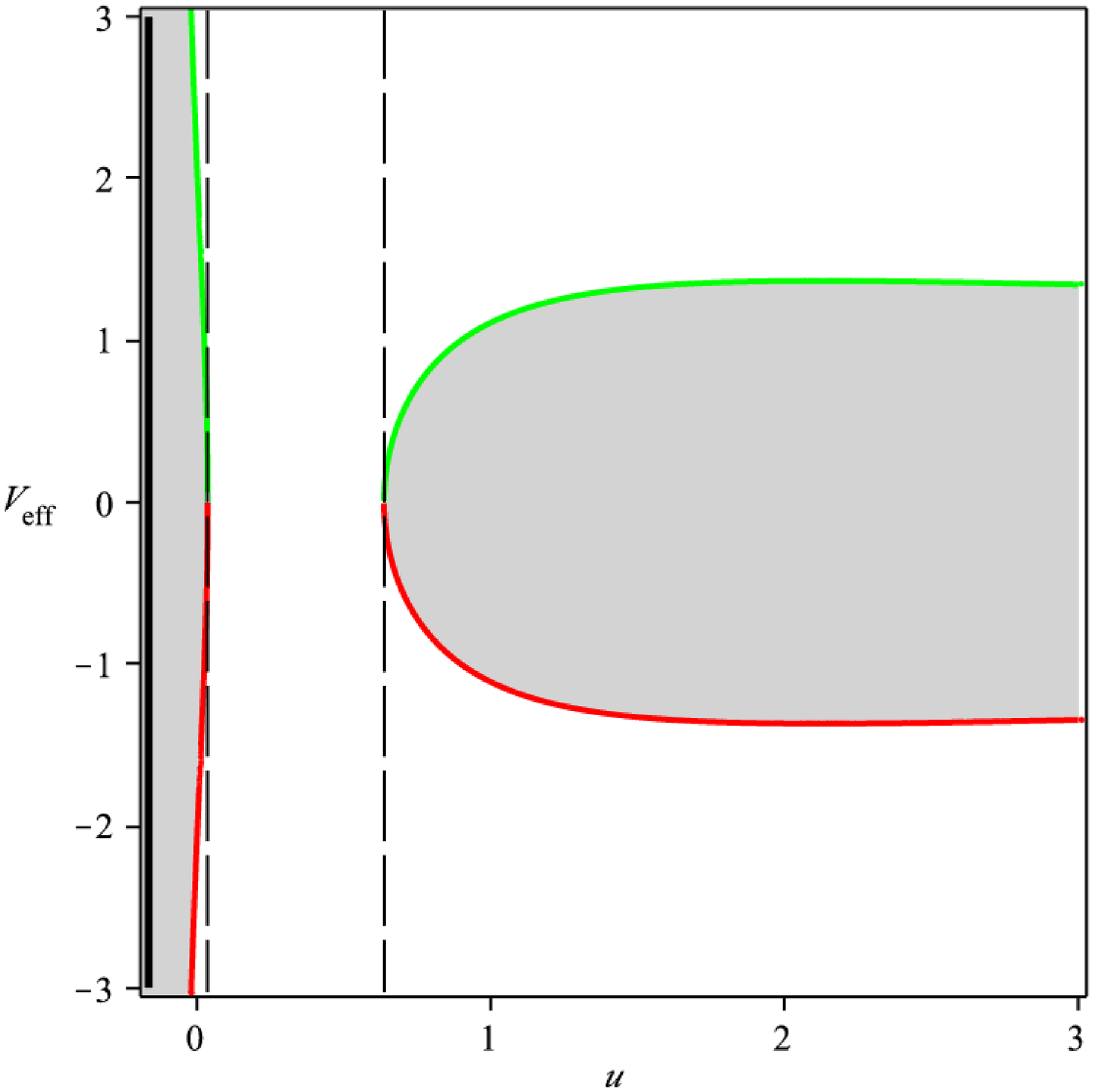}}
\subfigure[][$r_0=1, a=0.4, K=3, \Phi=0.9, \Psi=-0.9$. The dashed region is vorbidden.]{\label{pot2}\includegraphics[width=4.5cm]{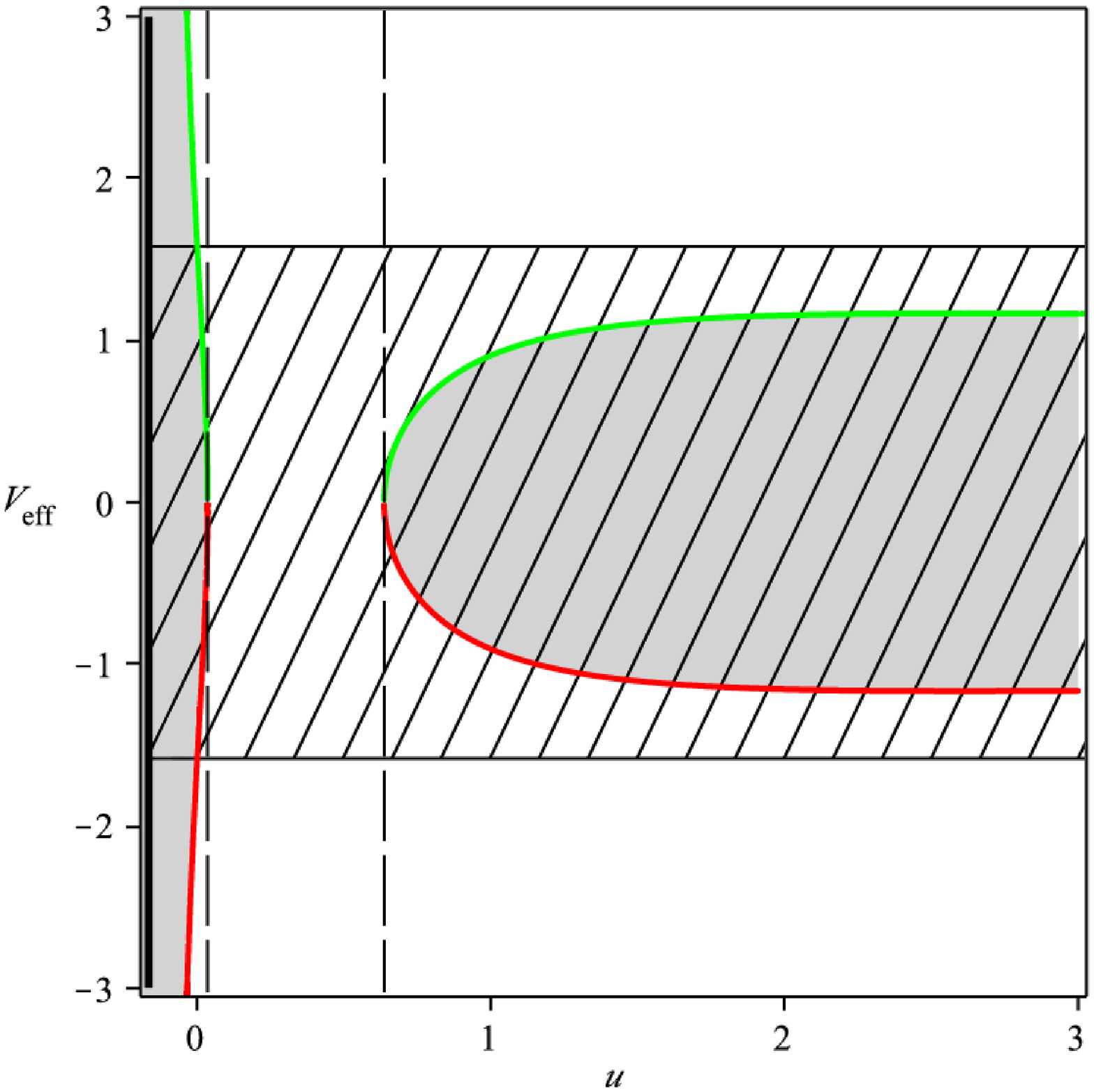}}
\subfigure[][$r_0=1, a=0.499, K=5, \Phi=-1, \Psi=1$. Nearly extreme Myers-Perry space-time.]{\label{pot3}\includegraphics[width=4.5cm]{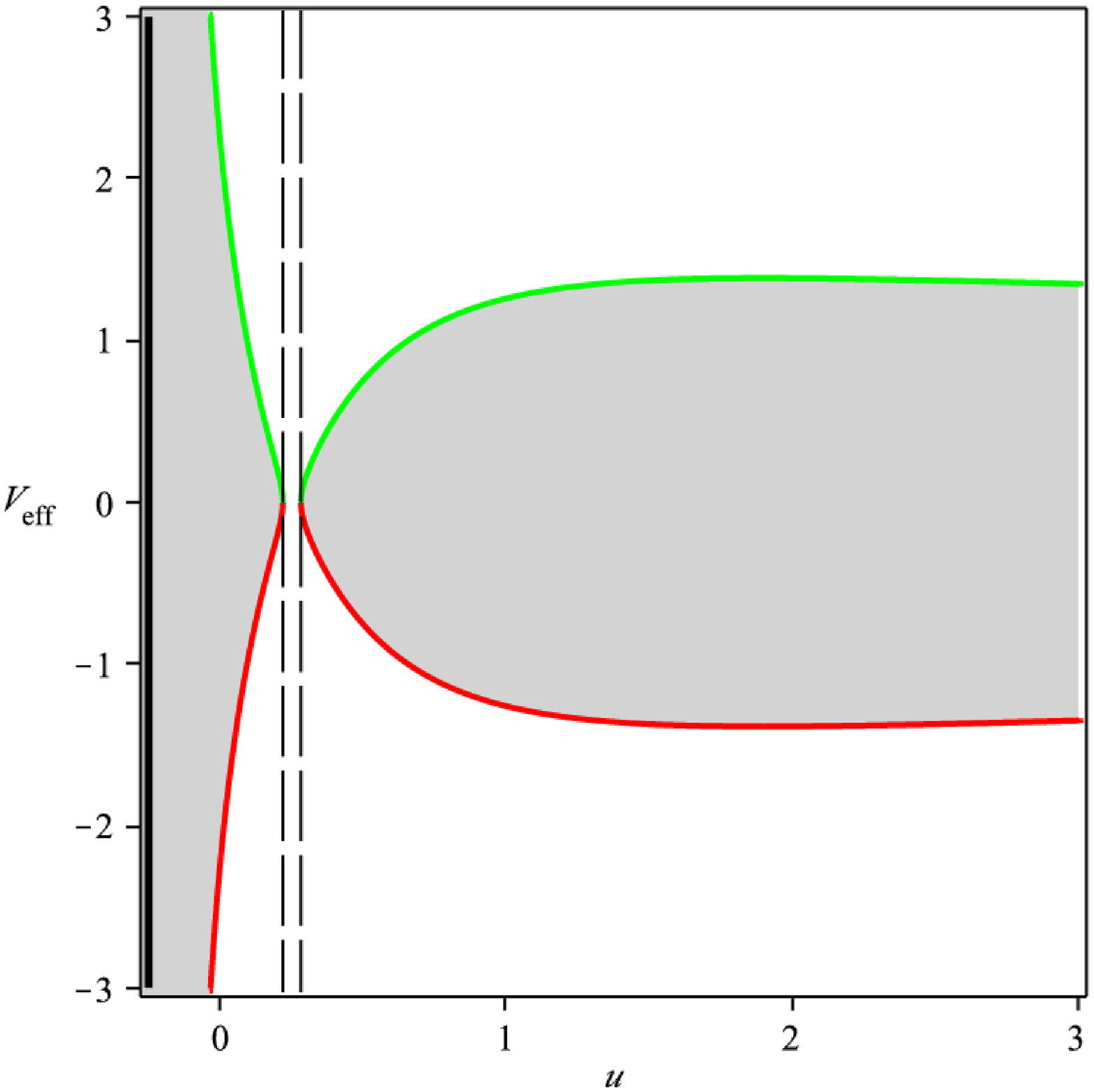}} \\
\subfigure[][$r_0=1, a=0.4, K=5, \Phi=-1, \Psi=-1$. Illustration of the types of orbits A, B and C from the Table~\ref{tab:orbits}.]{\label{pot4}\includegraphics[width=4.5cm]{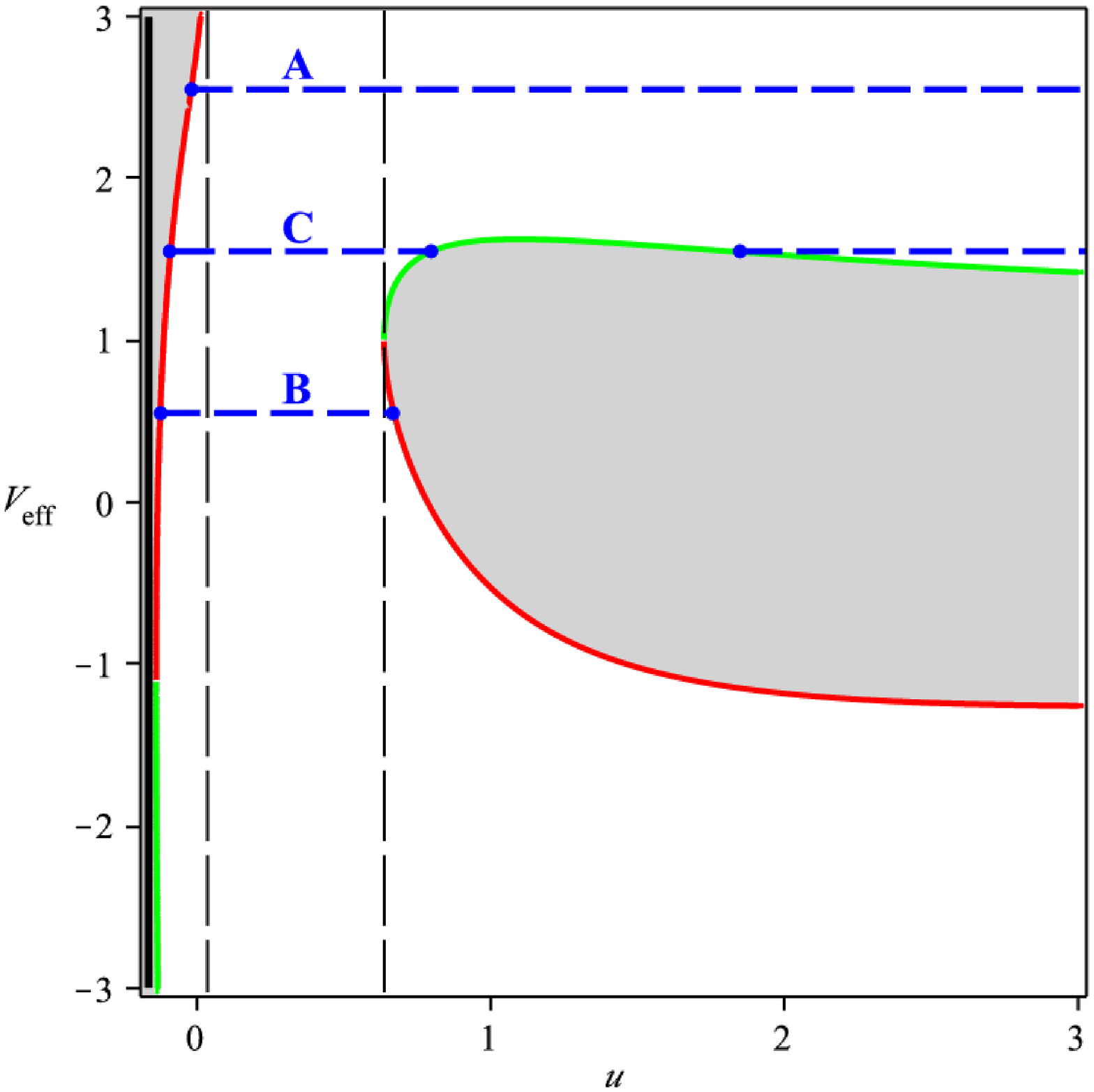}}
\subfigure[][$r_0=1, a=0.4, K=10, \Phi=-4, \Psi=-0.5$. The parameters are chosen from the region (4) in the Fig.~\ref{x0f}.]{\label{pot5}\includegraphics[width=4.5cm]{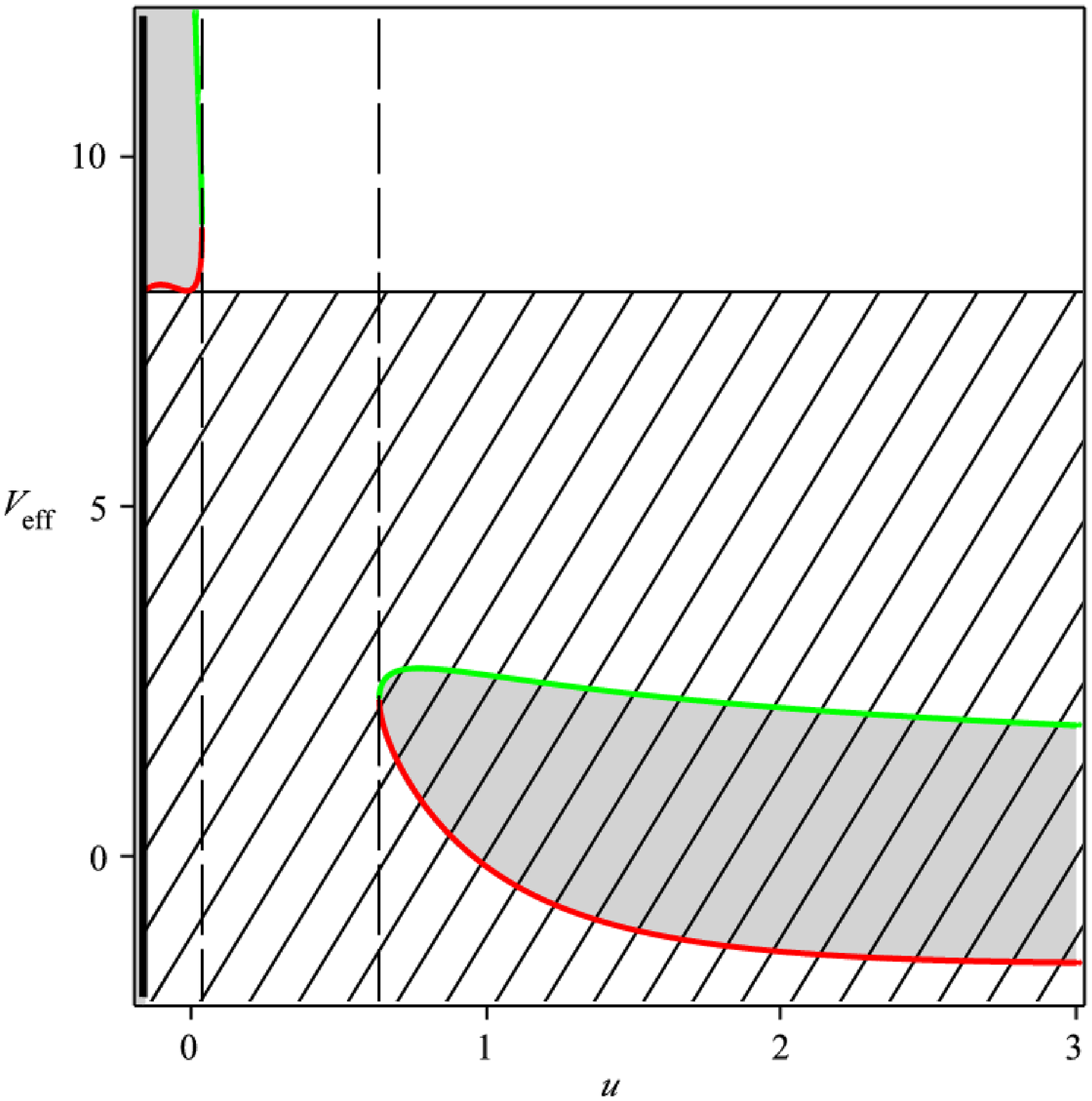}}
\subfigure[][$r_0=1, a=0.4, K=10, \Phi=-4, \Psi=-0.5$. Detailed representation of~\subref{pot5}, BO.]{\label{pot6}\includegraphics[width=4.5cm]{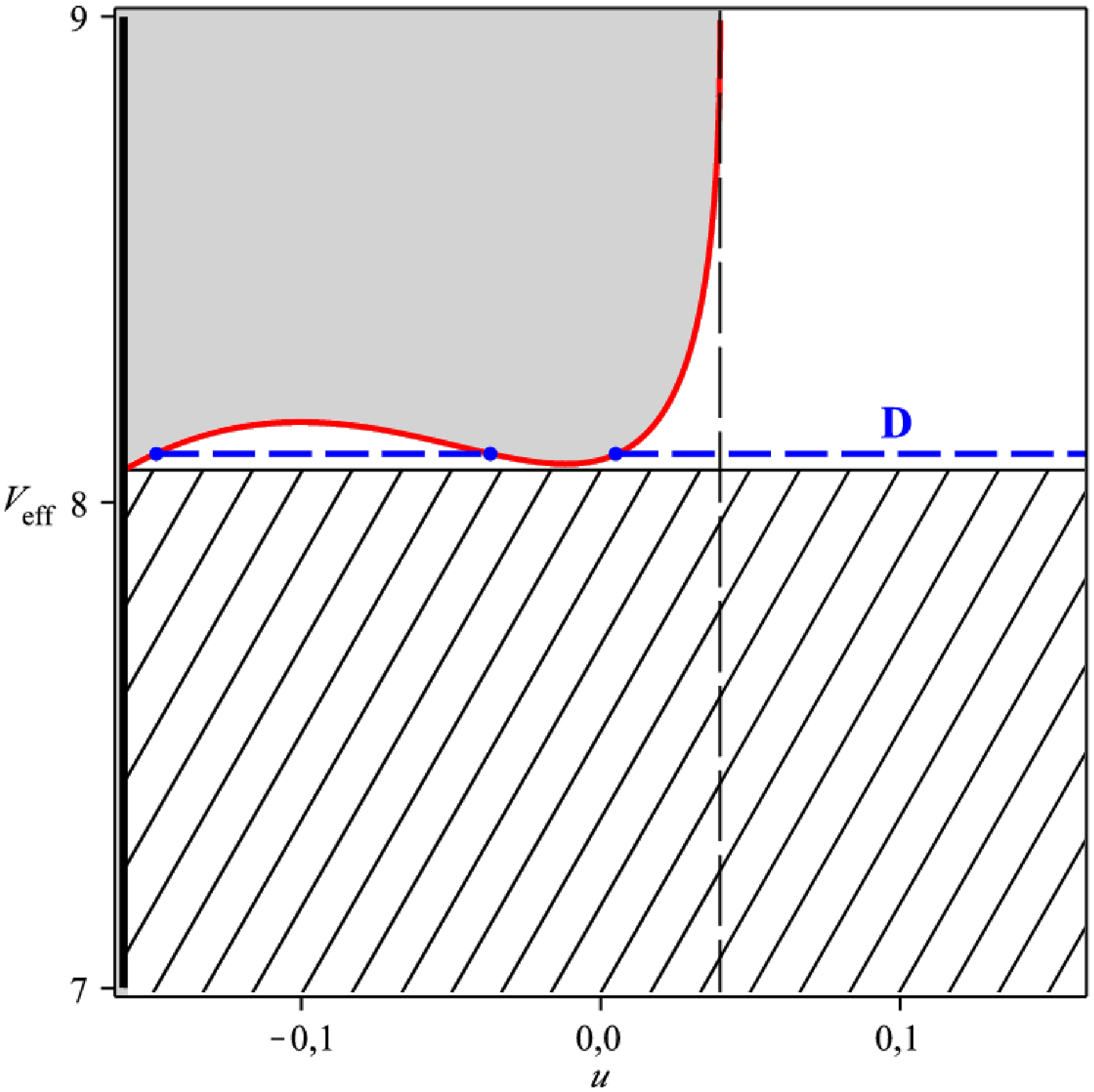}}
\end{center}
\caption{Effective potential for massive test particles ($\delta = 1$). The potentials $V_{\text{eff}}^+$ (red) and $V_{\text{eff}}^-$ (green) couple at the horizons (dashed lines). In the grey regions no motion is possible. The dashed region is forbidden from the conditions on the $\vartheta$-motion defined by the inequalities~\eqref{theta-cond-lamu}. The bold black vertical line at $u=-a^2$ defines the singularity. The dashed lines indicate the horizons $u_+$ and $u_-$. 
\label{fig:potentiale} }
\end{figure*}

\begin{figure*}[th!]
\begin{center}
\subfigure[][$r_0=1, a=0.4, K=5, \Phi=-0.5, \Psi=0.5$. Since $\Phi + \Psi = 0$ the plot is symmetric wrt the u-axis]{\label{pot1l}\includegraphics[width=4.5cm]{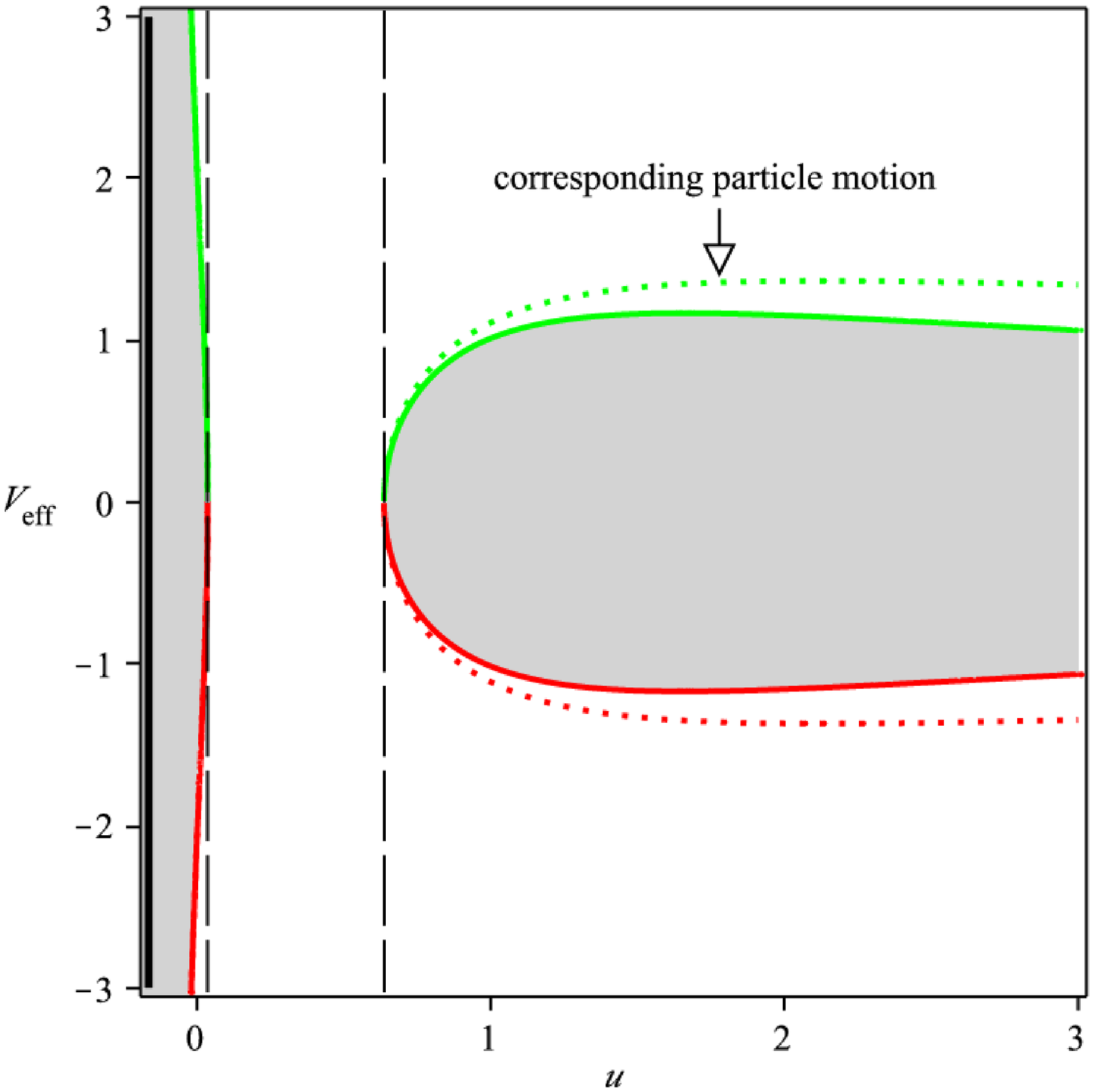}}
\subfigure[][$r_0=1, a=0.499, K=3, \Phi=0.5, \Psi=-0.5$. Nearly extreme Myers-Perry space-time.]{\label{pot3l}\includegraphics[width=4.5cm]{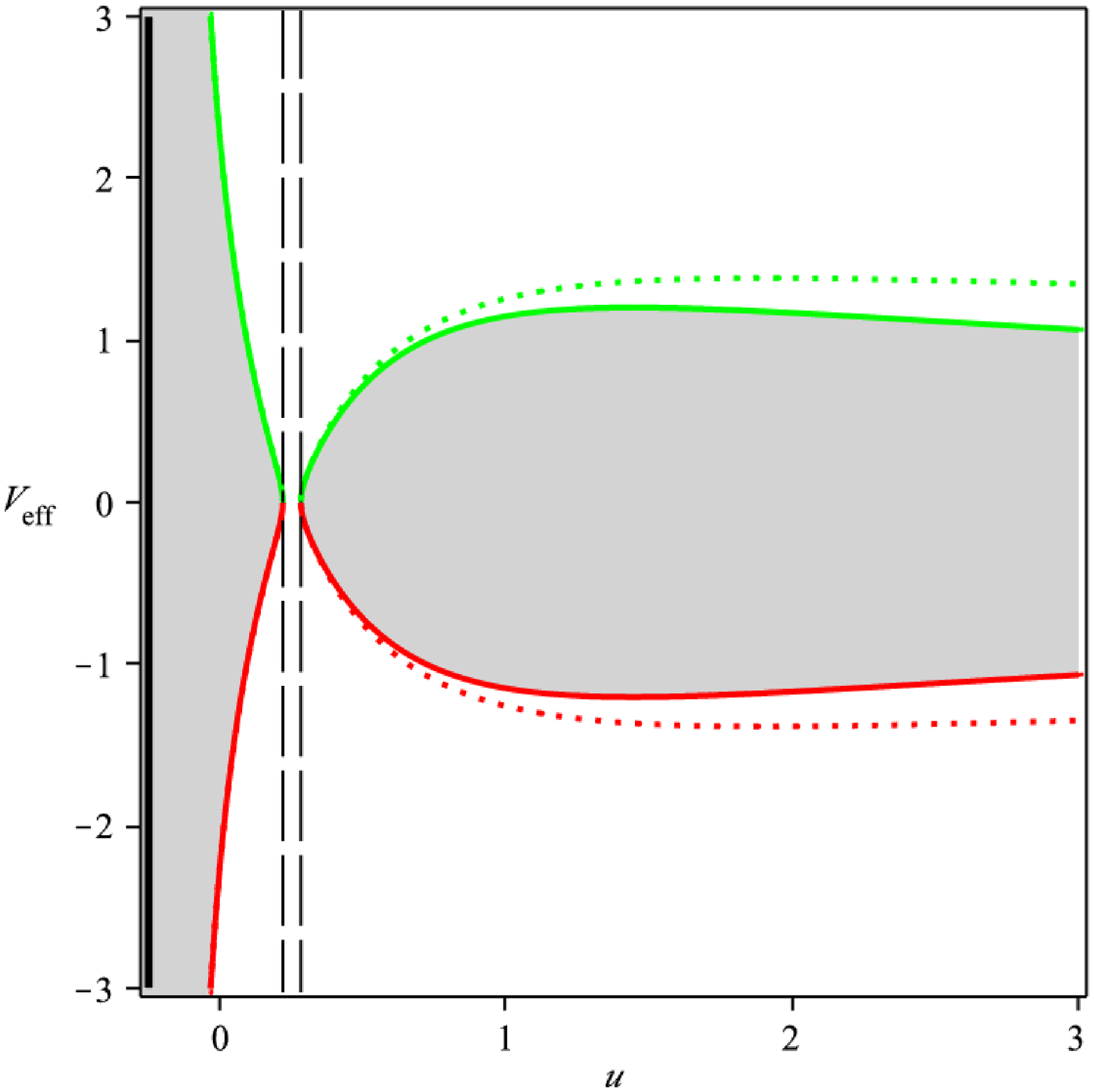}} 
\subfigure[][$r_0=1, a=0.4, K=5, \Phi=-1, \Psi=-1$.]{\label{pot4l}\includegraphics[width=4.5cm]{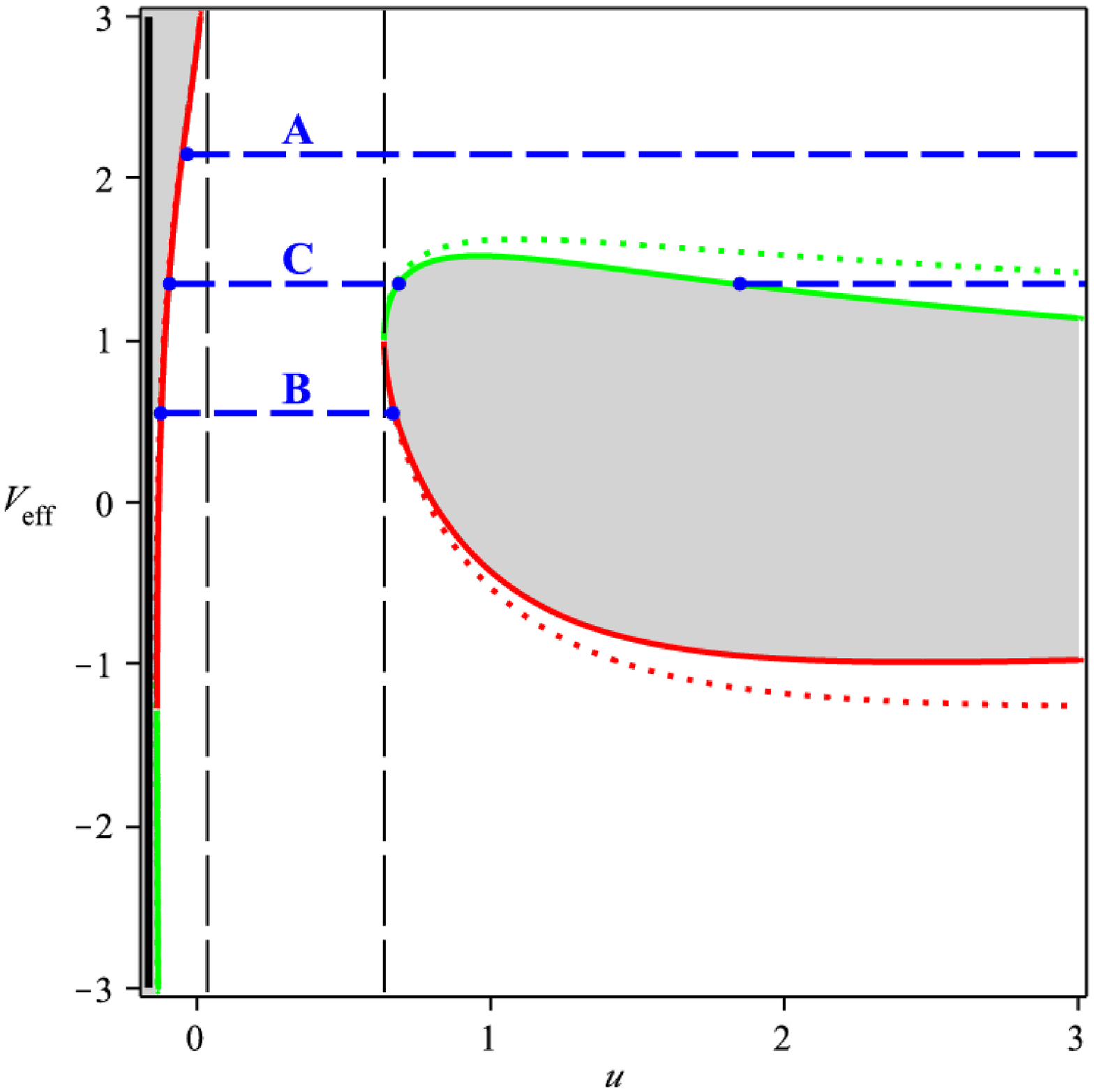}}
\subfigure[][$r_0=1, a=0.4, K=10, \Phi=-4, \Psi=-0.5$.]{\label{pot5l}\includegraphics[width=4.5cm]{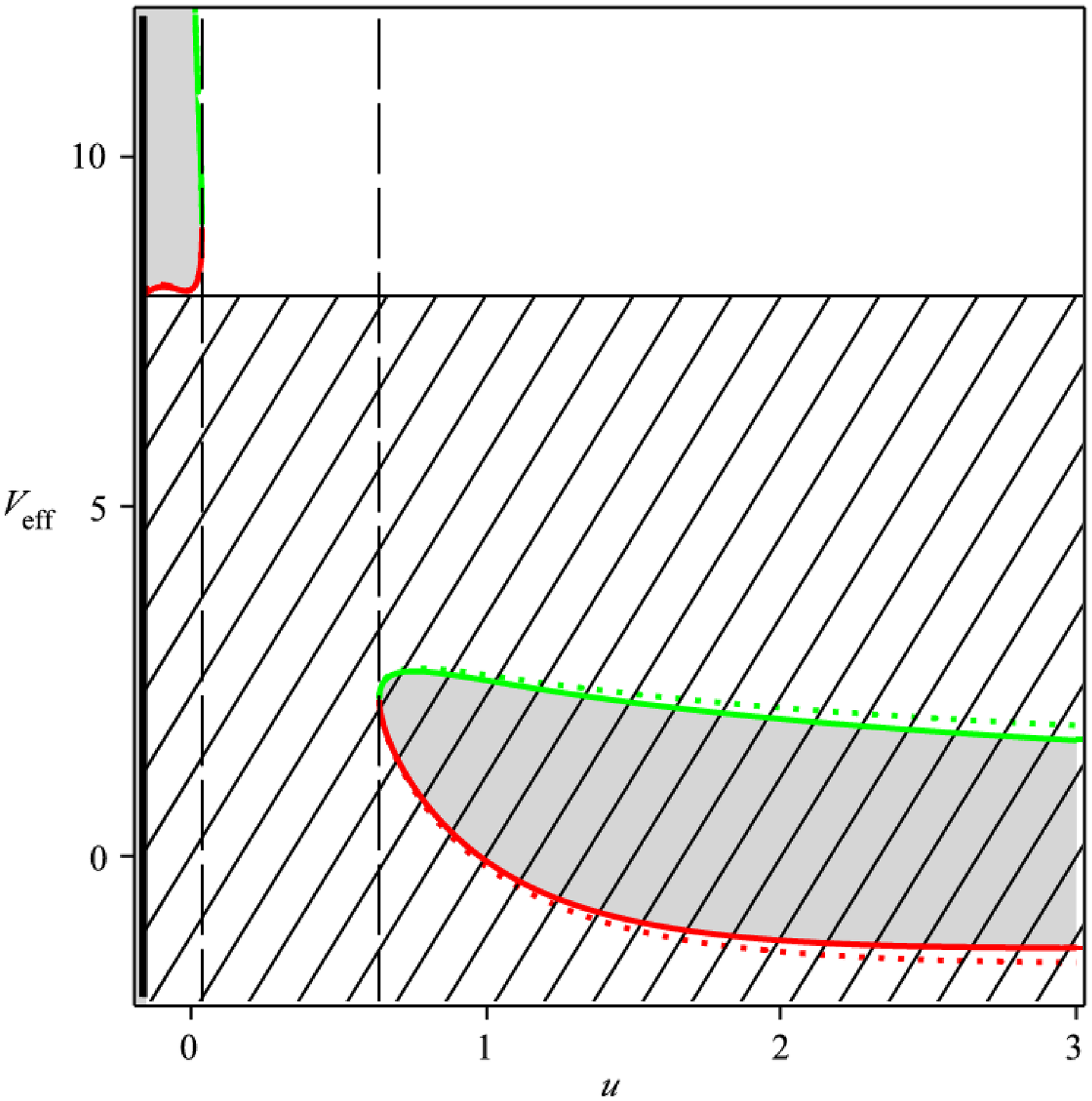}}
\subfigure[][$r_0=1, a=0.4, K=10, \Phi=-4, \Psi=-0.5$. Detailed representation of~\subref{pot5l}, BO.]{\label{pot6l}\includegraphics[width=4.5cm]{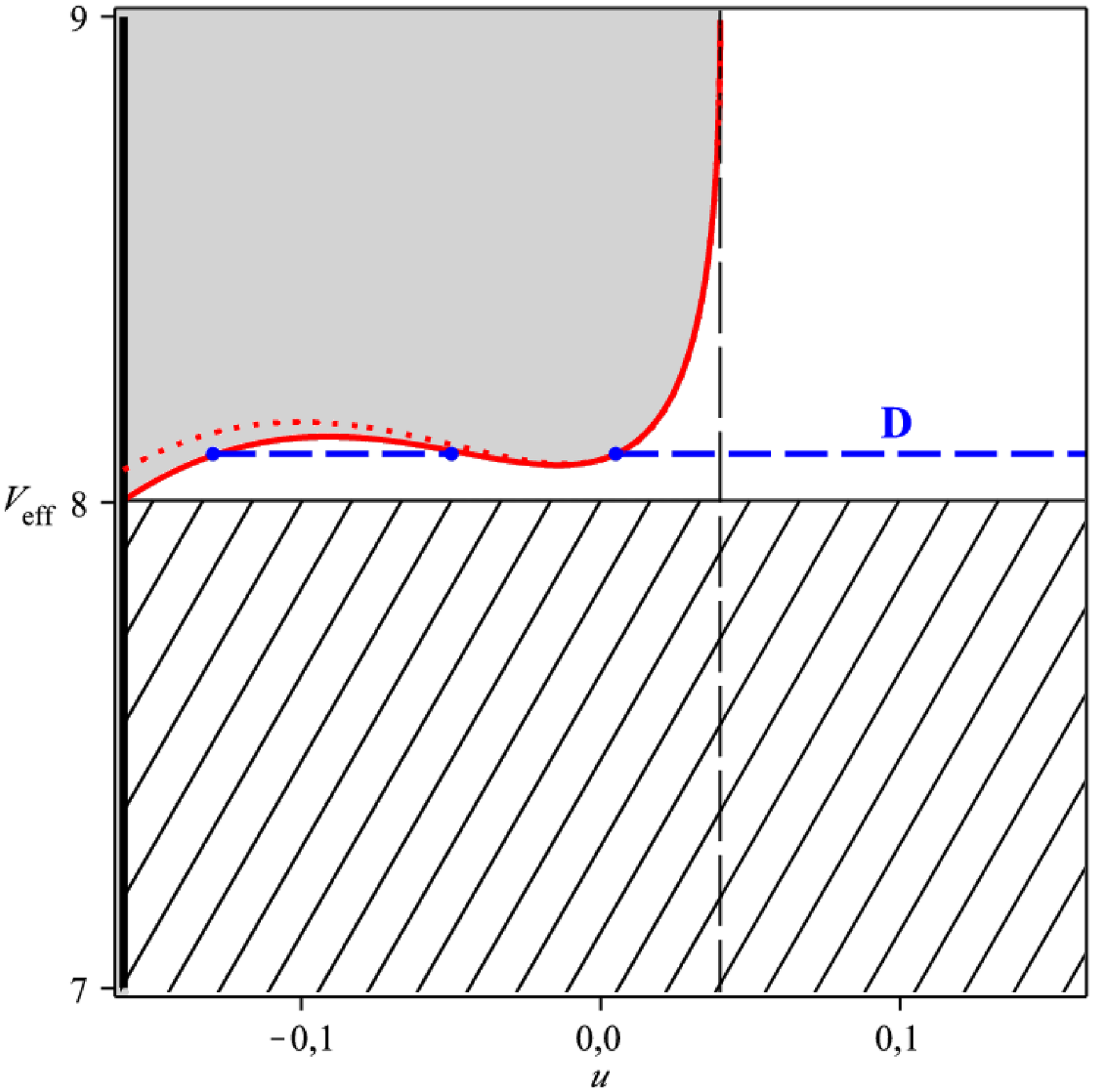}}
\end{center}
\caption{Effective potential for massless test particles ($\delta = 0$). The dotted lines in the potential are plotted for comparison and correspond to the test particle with non zero mass. One can see that the types of orbits from Table~\ref{tab:orbits} are conserved for light. \newline
The potentials $V_{\text{eff}}^+$ (red) and $V_{\text{eff}}^-$ (green) couple at the horizons (dashed lines). In the grey regions no motion is possible. The dashed region is forbidden from the conditions on the $\vartheta$-motion defined by the inequalities~\eqref{theta-cond-lamu}. The bold black vertical line at $u=-a^2$ defines the singularity. The dashed lines indicate the horizons $u_+$ and $u_-$. 
\label{fig:potentiale_licht} }
\end{figure*}

The potentials are shown in Fig.~\ref{fig:potentiale} for massive and in Fig.~\ref{fig:potentiale_licht} for massless test particles. For light the same types of orbits as for the massive test particles exist. The motion is forbidden in the grey region. At infinity $\lim\limits_{u \rightarrow \infty}{V_{\text{eff}}^{\pm}} = \pm \sqrt{\delta}$ is true. The potentials $V_{\text{eff}}^{+}$ and $V_{\text{eff}}^{-}$ meet at the horizons where $\Delta = 0$: 
\begin{eqnarray}
&& V_{\text{eff}}^{\pm} (u_{+}) = -\frac{a}{u_{+}+a^2} (\Phi + \Psi) \ , \\  \nonumber
&& V_{\text{eff}}^{\pm} (u_{-}) = -\frac{a}{u_{-}+a^2} (\Phi + \Psi) \ .
\end{eqnarray}
If $\Phi+\Psi=0$ then the ordinates of both points where the potentials meet lie at the $u$-axis and the potentials are symmetric wrt the $u$-axis: $V_{\text{eff}}^+$ = -$V_{\text{eff}}^-$ (Fig.~\ref{pot1}). The potentials are also symmetric wrt the change of sign of the parameters of the test particle: $V_{\text{eff}}^\pm \left(E, \Phi, \Psi \right)$ =  $V_{\text{eff}}^\pm \left(-E, -\Phi, -\Psi \right)$. 


There is in general a potential barrier which prohibits a test particle from falling into the singularity (Figs.~\ref{fig:potentiale}\subref{pot1}-\subref{pot4}). The potential plots are truncated at the singularity. If a part of a potential is truncated at the singularity, the remaining part containing no potential barrier is not allowed by the inequalities~\eqref{theta-cond-lamu} as in the plots~\subref{pot5} and \subref{pot6}. But as we know from the discussion above it is possible that a turning point of an orbit coincides with the singularity which implies that the orbit ends in the singularity. 

Setting $u=-a^2$ in the polynomial $U$ with the coefficients~\eqref{eq:33} one sees that only for $c_2=0$ the singularity can be reached: $U(u=-a^2)=-4a^2r_0^2c_2$. Thus, if we fix $\Psi$ and $\Phi$ then only particles with energy 
\begin{equation}
E_{c_2} = \pm \frac{1}{a} \sqrt{a^2 \delta - K + \left(\Phi + \Psi \right)^2} \, \label{Ec2}
\end{equation}
can fall into singularity. The orbits of these particles are denoted by the index $\rm s$ in the Table~\ref{tab:orbits}.

It is easy to see that only in this case falling into the singularity is possible. The potential is truncated at the singularity $u=-a^2$. At this point it takes the value
\begin{equation}
V_{\text{eff}}^{\pm} (-a^2) = \pm \frac{1}{a} \sqrt{ a^2 \delta - K + \left(\Phi + \Psi \right)^2 } \,
\end{equation} 
which coincides with~\eqref{Ec2}. Since $c_2\geq 0$ from the inequalities~\eqref{theta-cond-lamu}, the region between $E_{c_{2_{+}}}=V_{\text{eff}}^{+} (-a^2)$ and $E_{c_{2_{-}}}=V_{\text{eff}}^{-} (-a^2)$ is not allowed. But the energies $E_{c_2}$ corresponding to the boundary of this region where $c_2=0$ are allowed.

\section{Solution of the geodesic equation}\label{sec:solution}

Now we present the analytical solutions of the differential equations ~\eqref{eq-r-theta:1}--\eqref{dtdgamma}. 

\subsection{Solution of the $\vartheta$--equation}\label{vartheta-sol}

The solution of the differential equation~\eqref{eq-r-theta:2} is given by the elementary function
\begin{equation}
\vartheta(\gamma )=\arccos\Bigl( \pm \sqrt{ \frac{1}{2a_\xi}\left(  \sqrt{D}\sin\left( -\sqrt{-a_\xi}\gamma + \gamma^\vartheta_{\rm in}   \right) -b_\xi  \right)} \Bigr) \ ,
\end{equation}
where $a_\xi<0$ and $D>0$ are defined in~\eqref{xieom}, $\gamma ^\vartheta_{\rm in}=\sqrt{-a_\xi}\gamma _{{\rm in}} + \arcsin \left( \frac{2a_\xi\xi_{{\rm in}}+b_\xi}{\sqrt{b_\xi^2-4a_\xi c_\xi}}\right)$, $\gamma_{\rm in}$ is the initial value of $\gamma$ and $\xi_{\rm in}$ is the initial value of $\xi$.

\subsection{Solution of the $u$--equation}

A substitution $u=\frac{1}{b_3}\left(4y-\frac{b_2}{3}\right)$ transforms the polynomial $U$ into the standard Weierstra{\ss} form
\begin{equation}
\left(\frac{dy}{d\gamma}\right)^2=4y^3-g_2y-g_3:=P_3(y) \, ,  \label{P3}
\end{equation}
where 
\begin{equation}
g_2=\frac{b_2^2}{12} - \frac{b_1b_3}{4} \, , \qquad  g_3=\frac{b_1b_2b_3}{48} - \frac{b_0b_3^2}{16}-\frac{b_2^3}{216} \ .
\end{equation}
The differential equation \eqref{P3} is of elliptic type and is solved by the Weierstra{\ss}' $\wp$--function~\cite{Markush}
\begin{equation}
y(\gamma) = \wp\left(\gamma - \gamma'_{\rm in}; g_2, g_3\right) \ , \label{soly}
\end{equation}
where $\gamma'_{\rm in}=\gamma_{\rm in}+\int^\infty_{y_{\rm in}}{\frac{dy}{\sqrt{4y^3-g_2y-g_3}}}$
with $y_{\rm in}=\frac{b_3}{4} u_{\rm in} + \frac{b_2}{12}$.
Then the solution of~\eqref{eq-r-theta:1} acquires the form
\begin{equation}
u= \frac{1}{b_3}\left(4 \wp\left(\gamma - \gamma'_{\rm in}; g_2, g_3\right) -\frac{b_2}{3}\right)  \ . \label{solrNUTlight}
\end{equation}

\subsection{Solution of the $\varphi $--equation}

Eq.~\eqref{dvarphidgamma} can be simplified by using \eqref{eq-r-theta:1} and~\eqref{eq-r-theta:2}
and by performing the substitution $\xi = \cos^2\vartheta$ and $u=\frac{1}{b_3}\left( 4y-\frac{b_2}{3} \right)$
\begin{equation}
d\varphi = d\varphi_{\xi} + d\varphi_{u} \label{phi_new} \, ,
\end{equation}
where 
\begin{equation}
d{\varphi}_{\xi}=- \frac{d\xi}{\sqrt{\Theta_\xi}} \frac{{\Phi}}{1 - \xi} \,
\end{equation}
where $\Theta_\xi$ is given in~\eqref{xieom} for $a=b$, and
\begin{equation}
d{\varphi}_{u}= \frac{dy}{\sqrt{P_3(y)}} \sum^2_{j=1}\frac{K_j}{y-p_j}  \ , \label{dvarphi_u}
\end{equation}
where the partial fractions decomposition procedure was applied. Here $K_i$, $i=1,2$ are constants which arise from the partial fractions decomposition. These depend on the parameters of the metric and the test particle, and $p_1$ and $p_2$ are the zeros of $\Delta\equiv\Delta\left(u=\frac{1}{b_3}\left( 4y-\frac{b_2}{3} \right)\right)$.

The integration of~\eqref{phi_new} yields:
\begin{equation}
\varphi - \varphi_{\rm{in}} = \varphi_{\xi} + \varphi_{u} \label{phi_new2} \, ,
\end{equation}
where $\varphi_{\rm{in}}$ is the initial value of the angle $\varphi$.

Consider first the $d\varphi_{u}$. After the substitution $y=\wp(v)$ with 
$\wp^\prime(v)=\sqrt{4 \wp^3(v)-g_2\wp(v)-g_3}$ and integration of the differential~\eqref{dvarphi_u} one gets:
\begin{equation}
\varphi_{u} = \int^v_{v_{\rm in}} \sum^2_{j=1}\frac{K_j}{\wp(v)-\wp(v_{\wp_j})} dv \ . \label{varphi_u}
\end{equation}  
Here $v=v(\gamma)=\gamma-\gamma^\prime_{\rm in}$ and $v_{\rm in}=v(\gamma_{\rm in})$. And $\wp(v_{\wp_j})=p_j \,, \, j=1,2$.

The solution of~\eqref{varphi_u} is given in terms of the elliptic $\sigma$ and $\zeta$ functions~\cite{GK10,KKHL10}:
\begin{equation}
\varphi_{u}  = 
\sum^2_{j=1}\frac{K_j}{\wp^\prime(v_{\wp_j})}
\Biggl( 2\zeta(v_{\wp_j})(v-v_{\rm in}) + \ln\frac{\sigma(v-v_{\wp_j})}{\sigma(v+v_{\wp_j})}
- \ln\frac{\sigma(v_{\rm in}-v_{\wp_j})}{\sigma(v_{\rm in}+v_{\wp_j})} \Biggr)   \label{varphi_u2}  \ ,
\end{equation}

The integral $\varphi_{\xi}$ can be easily found and the solution for $a_{\xi} < 0$ and $D > 0$ is given by
\begin{equation}
\varphi_\xi(\gamma) = \frac{1}{2} \arctan\frac{ \sqrt{D} - u_\xi (b_\xi + 2a_\xi)}{4\Phi\sqrt{-a_\xi}\sqrt{1-u_\xi^2}}  \Bigl|^{\xi(\gamma)}_{\xi_{{\rm in}}}  \label{sol2phi} \, ,
\end{equation}
where
\begin{equation}
u_\xi = \frac{2 a_\xi \xi +b_\xi}{\sqrt{D}} \ . \label{eq:uxi}
\end{equation}

\subsection{Solution of the $\psi$--equation}

Like for the coordinate $\varphi$ the Eq.~\eqref{dpsidgamma} can be simplified by using \eqref{eq-r-theta:1} and~\eqref{eq-r-theta:2}
and by performing the substitution $\xi = \cos^2\vartheta$ and $u=\frac{1}{b_3}\left( 4y-\frac{b_2}{3} \right)$
\begin{equation}
d\psi = d\psi_{\xi} + d\psi_{u} \label{psi_new} \, ,
\end{equation}
where 
\begin{equation}
d{\psi}_{\xi}=- \frac{d\xi}{\sqrt{\Theta_\xi}} \frac{{\Psi}}{\xi} \,
\end{equation}
where $\Theta_\xi$ is given in~\eqref{xieom} for $a=b$. And
\begin{equation}
d{\psi}_{u}= \frac{dy}{\sqrt{P_3(y)}} \sum^2_{j=1}\frac{N_j}{y-p_j}  \ , \label{dpsi_u}
\end{equation}
where analogously the partial fractions decomposition procedure was applied. Here $N_i$, $i=1,2$, are constants which arise from the partial fractions decomposition. These depend on the parameters of the metric and the test particle. And $p_1$ and $p_2$ are the zeros of  $\Delta\equiv\Delta\left(u=\frac{1}{b_3}\left( 4y-\frac{b_2}{3} \right)\right)$.

The integration of~\eqref{psi_new} yields:
\begin{equation}
\psi - \psi_{\rm{in}} = \psi_{\xi} + \psi_{u} \label{psi_new2} \, ,
\end{equation}
where $\psi_{\rm{in}}$ is the initial value of the angle $\psi$.

Consider first the $d\psi_{u}$. After the substitution $y=\wp(v)$ with 
$\wp^\prime(v)=\sqrt{4 \wp^3(v)-g_2\wp(v)-g_3}$ and integration of the differential~\eqref{dpsi_u} one gets:
\begin{equation}
\psi_{u} = \int^v_{v_{\rm in}} \sum^2_{j=1}\frac{N_j}{\wp(v)-\wp(v_{\wp_j})} dv \ . \label{psi_u}
\end{equation}  
Here $v=v(\gamma)=\gamma-\gamma^\prime_{\rm in}$ and $v_{\rm in}=v(\gamma_{\rm in})$. And $\wp(v_{\wp_j})=p_j \,, \, j=1,2$.

The solution of~\eqref{psi_u} is given in terms of the elliptic $\sigma$ and $\zeta$ functions:
\begin{equation}
\psi_{u}  = 
\sum^2_{j=1}\frac{N_j}{\wp^\prime(v_{\wp_j})}
\Biggl( 2\zeta(v_{\wp_j})(v-v_{\rm in}) + \ln\frac{\sigma(v-v_{\wp_j})}{\sigma(v+v_{\wp_j})}
- \ln\frac{\sigma(v_{\rm in}-v_{\wp_j})}{\sigma(v_{\rm in}+v_{\wp_j})} \Biggr)   \label{psi_u2}  \ ,
\end{equation}

$d\psi_{\xi}$ can be easily integrated and the solution for $a_{\xi} < 0$ and $D > 0$ is given by
\begin{equation}
\psi_\xi(\gamma) = \frac{1}{2} \arctan{ \frac{b_\xi u_\xi - \sqrt{D} }{4\Psi\sqrt{-a_\xi} \sqrt{1-u_\xi^2}}}   \Bigl|^{\xi(\gamma)}_{\xi_{{\rm in}}} \label{sol2psi} \, ,
\end{equation}
where $u_\xi$ is defined in~\eqref{eq:uxi}.

\subsection{Solution of the $t$--equation}

Like for the coordinates $\varphi$ and $\psi$ the Eq.~\eqref{dtdgamma} can be simplified by using \eqref{eq-r-theta:1} and by performing the substitution $u=\frac{1}{b_3}\left( 4y-\frac{b_2}{3} \right)$ (for $a=b$ only the $u$-part survives)
\begin{equation}
d{t}_{u}= \frac{dy}{\sqrt{P_3(y)}} \left( \frac{4E}{b_3}y + T_0 + \sum^2_{j=1}\frac{T_j}{y-p_j} \right) \ , \label{dt_u}
\end{equation}
where the partial fractions decomposition procedure was applied. Here $p_2$ and $p_1$ are defined as before in the $\varphi$ and $\psi$ equations. $T_i$, $i=0,1,2$ are constants which arise from the partial fractions decomposition. These depend on the parameters of the metric 
and the test particle.

The integration of~\eqref{dt_u} yields:
\begin{equation}
t - t_{\rm{in}} = t_{u} \label{dt_new} \, ,
\end{equation}
where $t_{\rm{in}}$ is the initial value of $t$.

Substitution of $y=\wp(v)$ with 
$\wp^\prime(v)=\sqrt{4 \wp^3(v)-g_2\wp(v)-g_3}$ in~\eqref{dt_u} yields:
\begin{equation}
t_{u} =  \int^v_{v_{\rm in}} \left( \frac{4E}{b_3}\wp(v) + T_0 + \sum^2_{j=1}\frac{T_j}{\wp(v)-\wp(v_{\wp_j})} \right) dv  \label{t_u} \ .
\end{equation}  
Here $v=v(\gamma)=\gamma-\gamma^\prime_{\rm in}$ and $v_{\rm in}=v(\gamma_{\rm in})$, and $\wp(v_{\wp_j})=p_j \,, \, j=1,2$.

The solution of~\eqref{t_u} is given in terms of the elliptic $\sigma$ and $\zeta$ functions:
\begin{equation}
t_{u}  = \frac{4E}{3} \left(\zeta(v_{\rm in}) - \zeta(v)\right)  + T_0 (v-v_{\rm in}) + \sum^2_{j=1}\frac{T_j}{\wp^\prime(v_{\wp_j})}
\Biggl( 2\zeta(v_{\wp_j})(v-v_{\rm in}) + \ln\frac{\sigma(v-v_{\wp_j})}{\sigma(v+v_{\wp_j})}
- \ln\frac{\sigma(v_{\rm in}-v_{\wp_j})}{\sigma(v_{\rm in}+v_{\wp_j})} \Biggr)   \label{t_u2}  \ ,
\end{equation}

\section{The orbits}\label{sec:orbits}

\begin{figure*}[th!]
\begin{center}
\subfigure[][MBO. $r_0=1, a=0.2, K=3, E=1.14756215, \Phi=0.2, \Psi=0.2$.]{\label{orb1}\includegraphics[width=5.5cm]{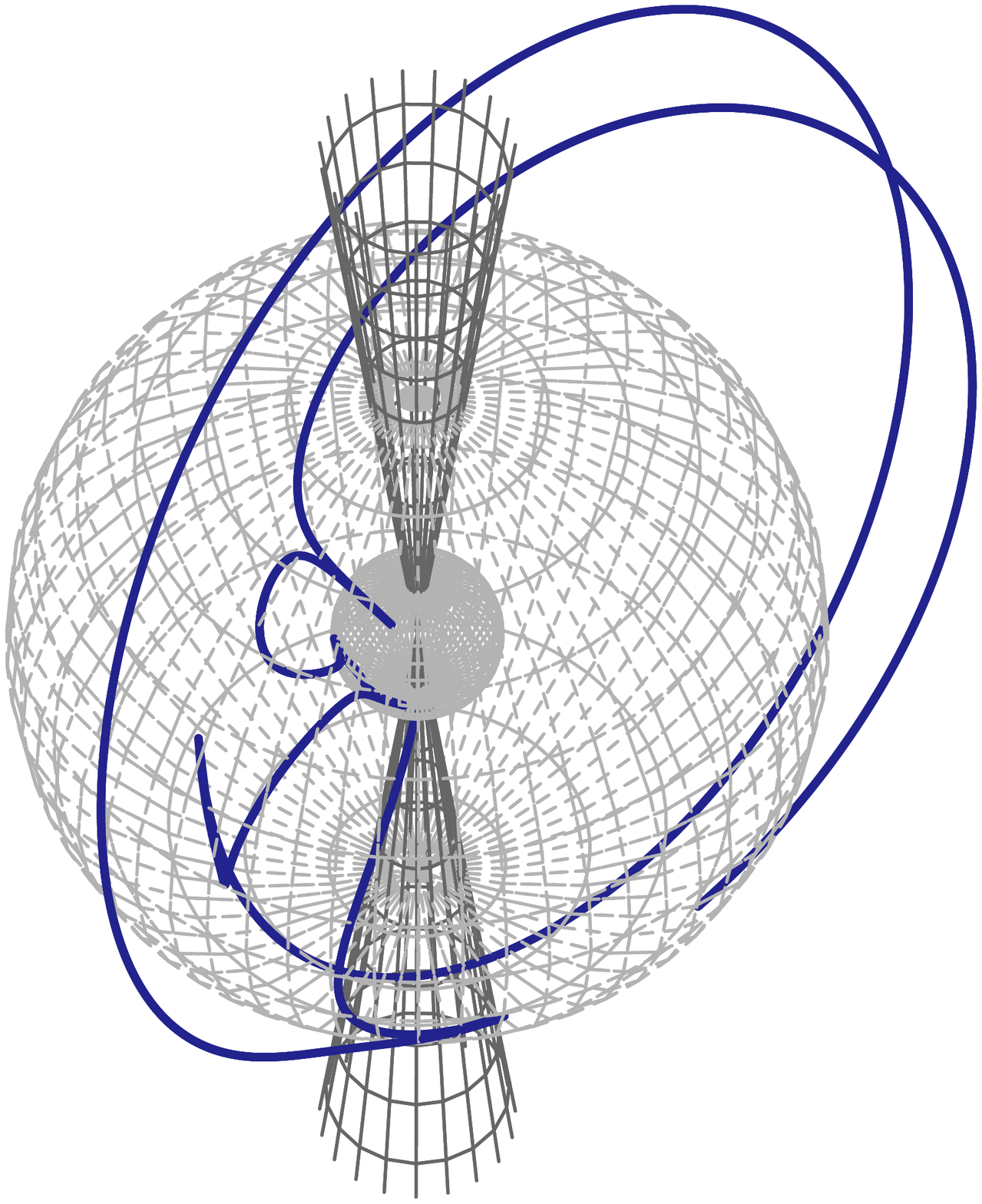}} \qquad
\subfigure[][EO. $r_0=1, a=0.2, K=3, E=1.14756215, \Phi=0.2, \Psi=0.2$. ]{\label{orb3}\includegraphics[width=5.5cm]{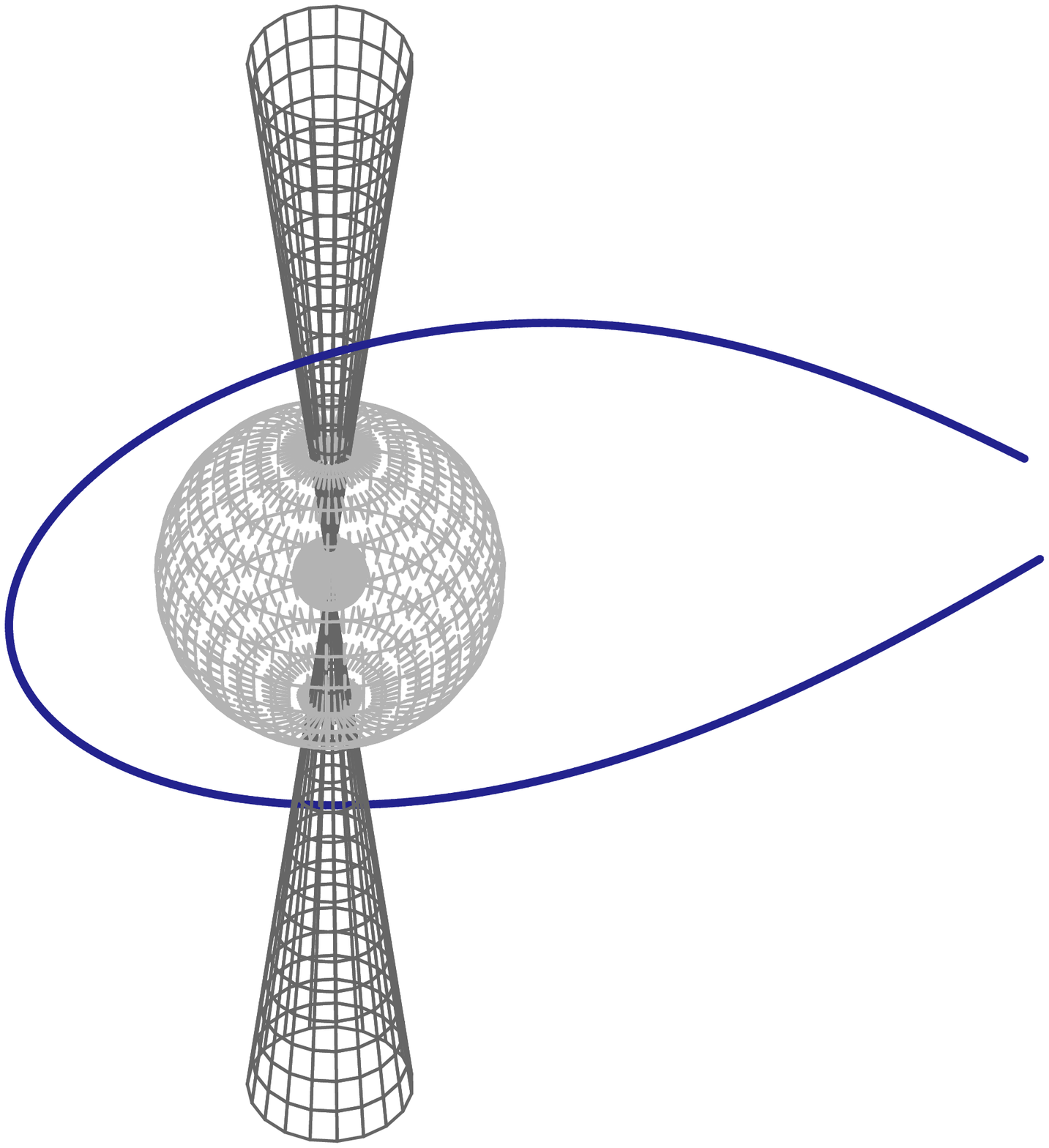}} \\ 
\subfigure[][BO. $r_0=1, a=0.4, K=10, E=8.16516573, \Phi=-4, \Psi=-0.5$. ]{\label{orb2}\includegraphics[width=5.5cm]{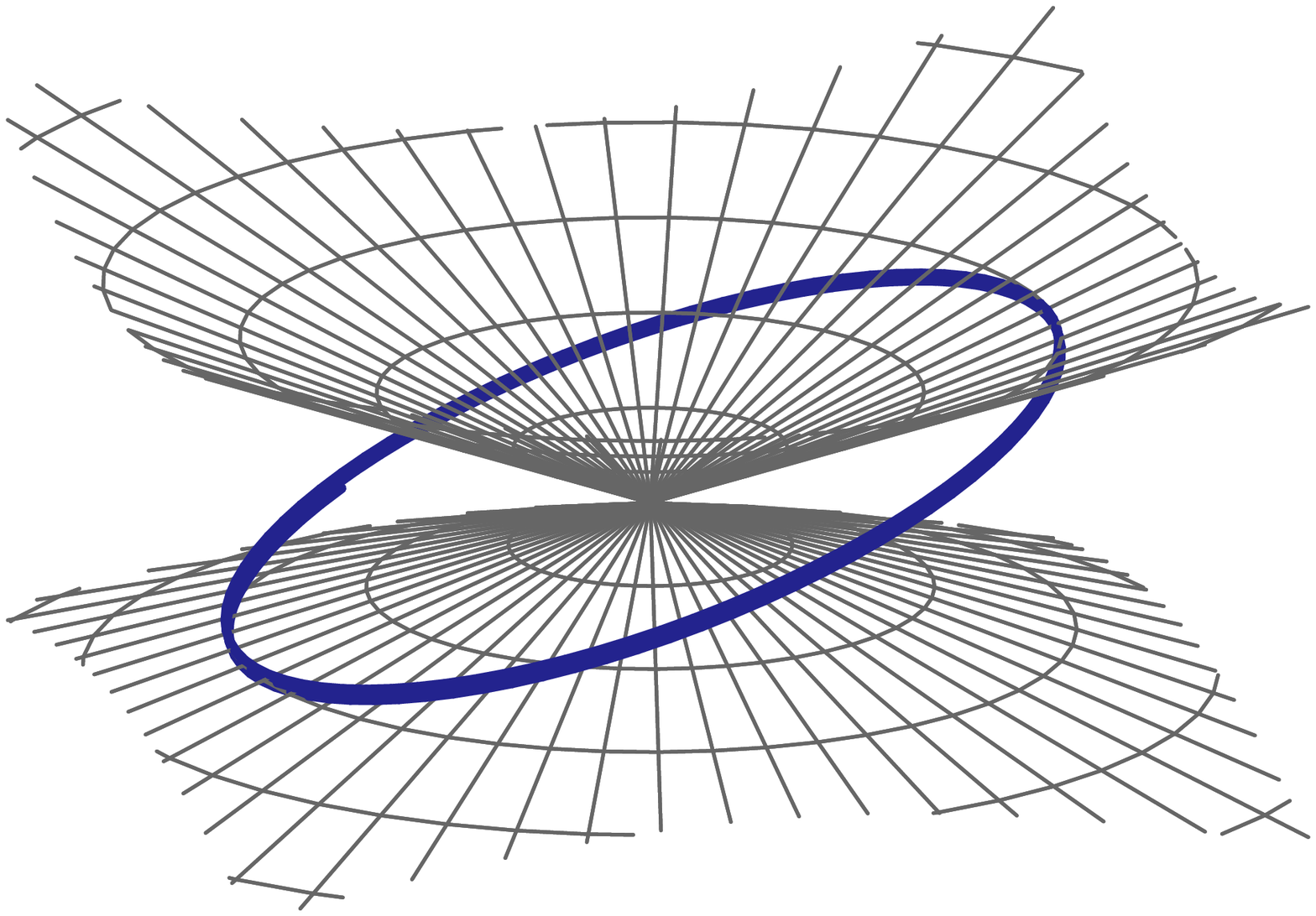}} \qquad 
\subfigure[][BO. $r_0=1, a=0.4, K=10, E=8.126604544, \Phi=-4, \Psi=-0.5$. ]{\label{orb2_2}\includegraphics[width=5.5cm]{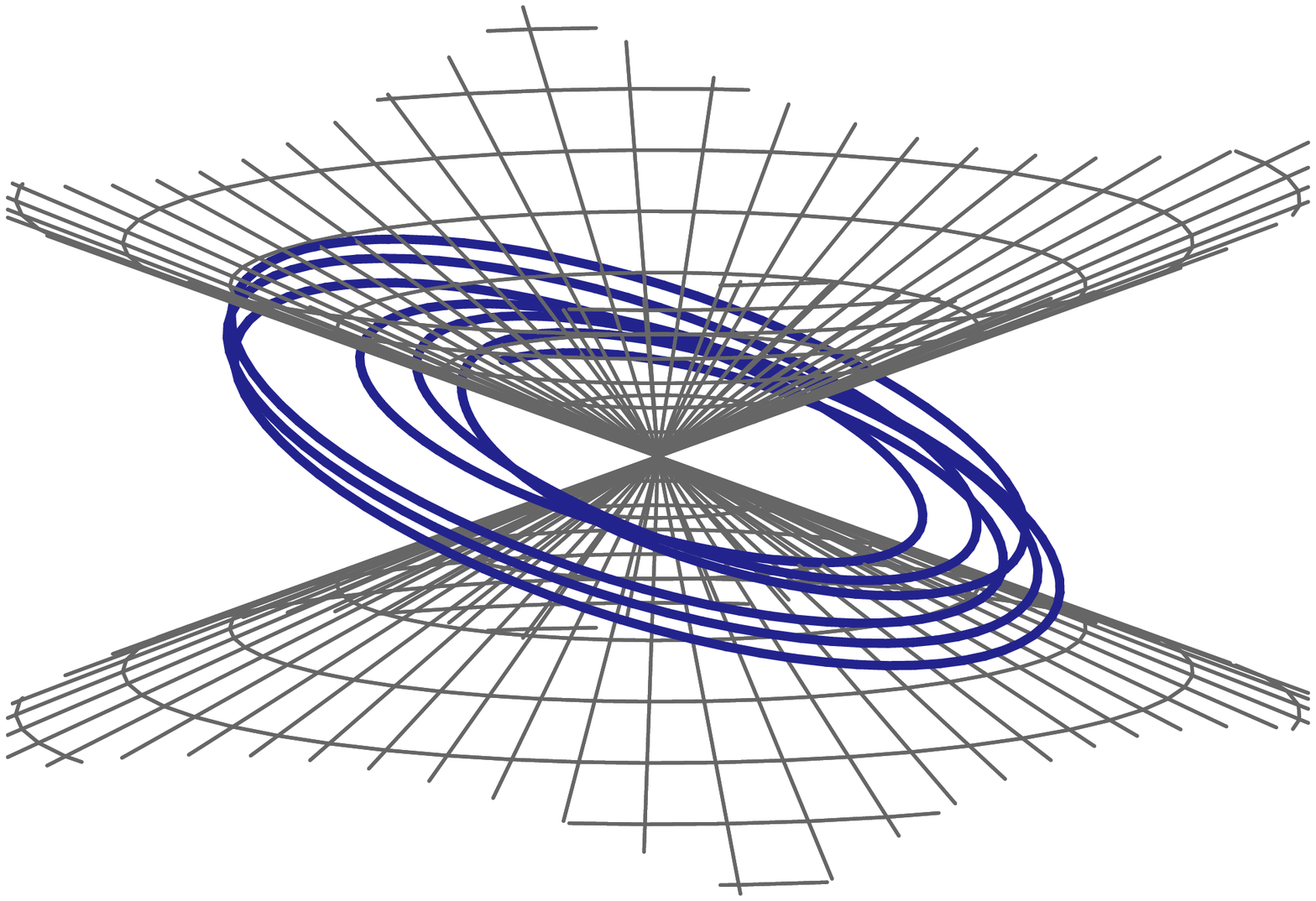}} \qquad 
\subfigure[][TEO. $r_0=1, a=0.4, K=10, E=8.126604544, \Phi=-4, \Psi=-0.5$. ]{\label{orb2_22}\includegraphics[width=5.0cm]{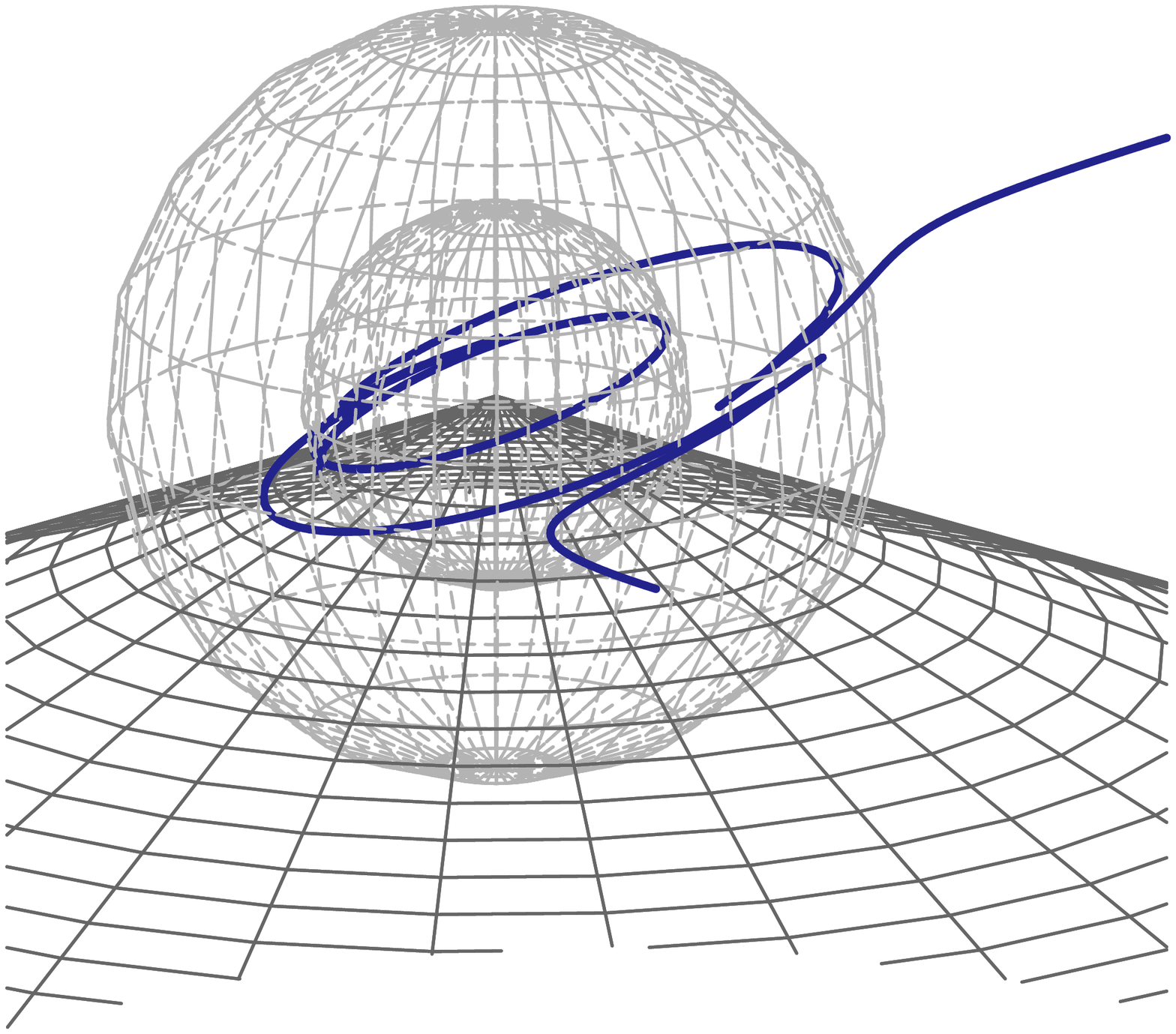}} \\
\subfigure[][TEO. $r_0=1, a=0.2, K=3, E=1.18247681, \Phi=-0.3, \Psi=0.9$.]{\label{orb4}\includegraphics[width=4.2cm]{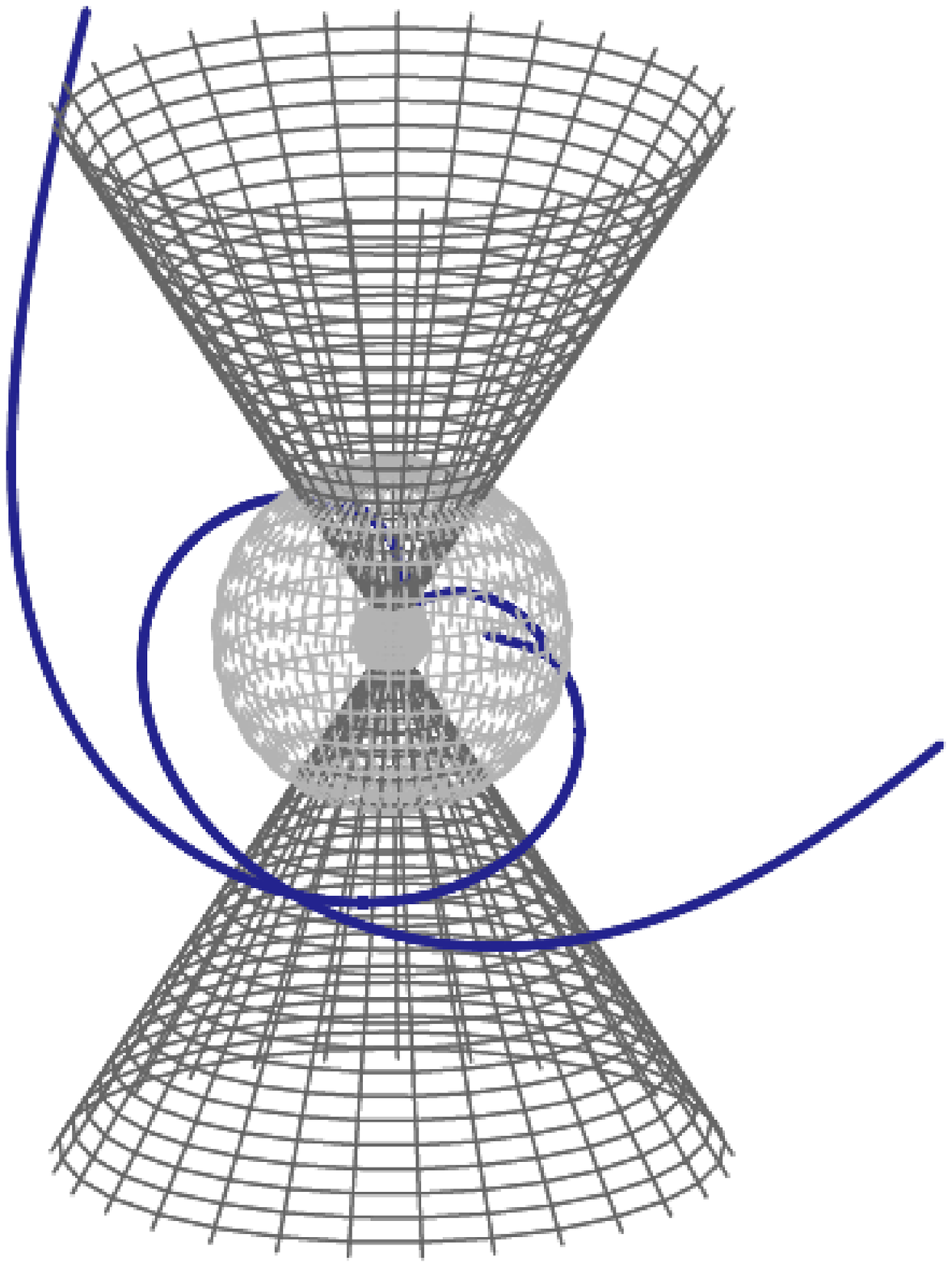}} \quad 
\subfigure[][EO. $r_0=1, a=0.4, K=5, E=1.625363954, \Phi=-1, \Psi=-1$.]{\label{orb5}\includegraphics[width=5.7cm]{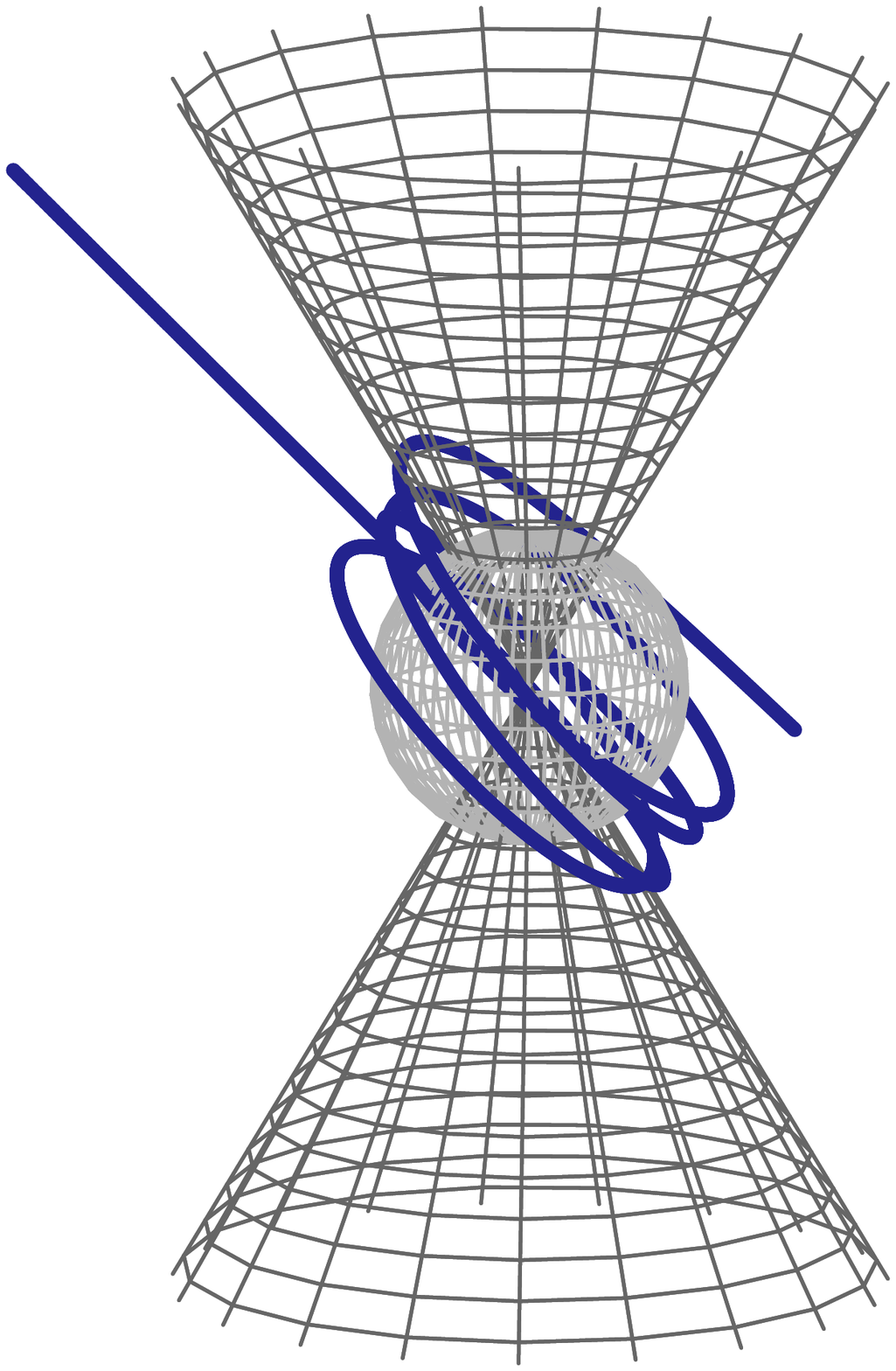}} \quad
\subfigure[][MBO. $r_0=1, a=0.4, K=5, E=1.625363954, \Phi=-1, \Psi=-1$.]{\label{orb6}\includegraphics[width=5.5cm]{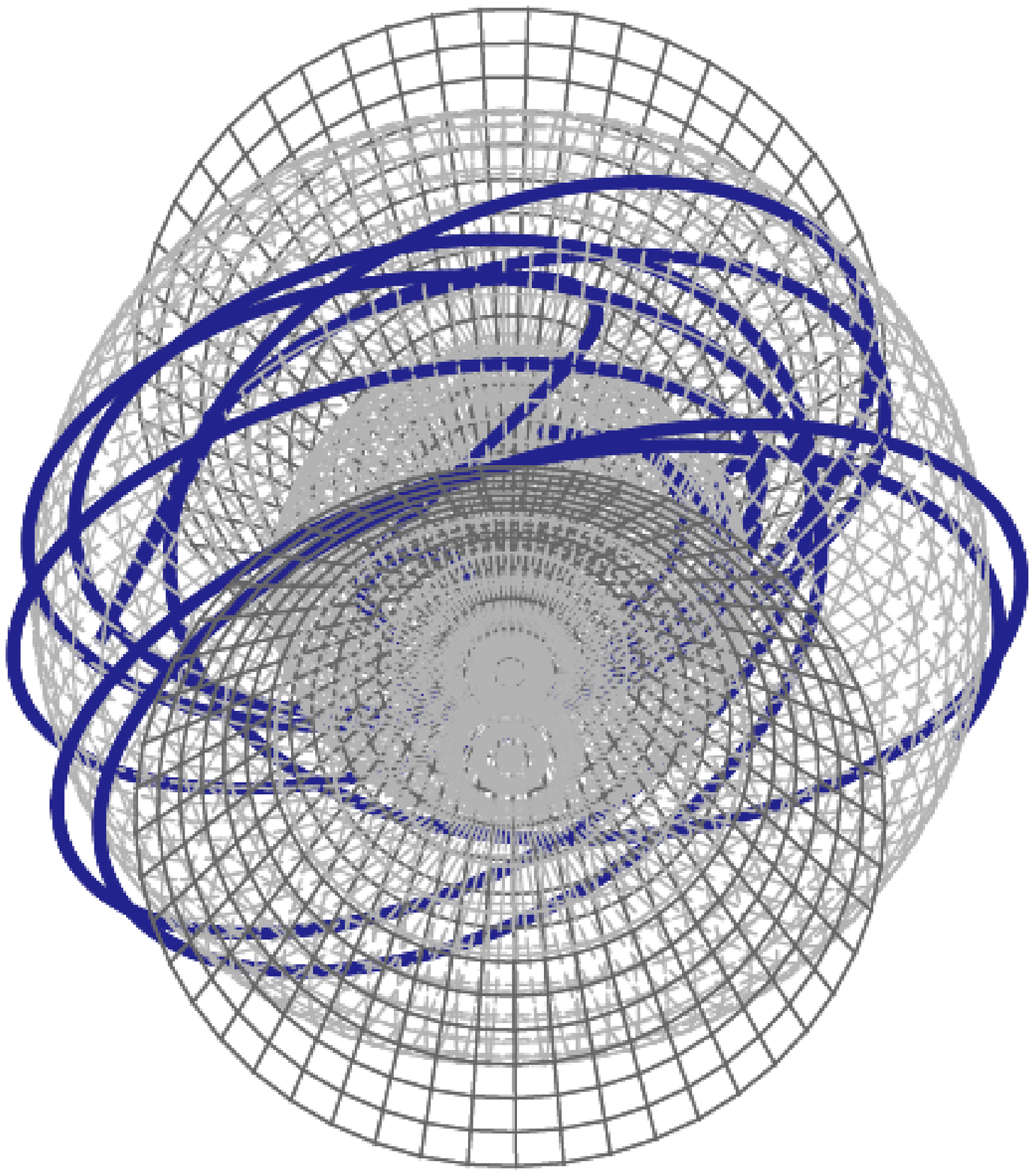}}
\end{center}
\caption{Orbits for $\delta=1$ corresponding to the possible types specified in the Table~\ref{tab:orbits}: MBO, BO, EO and TEO. In the picture~\subref{orb2_22} we omit the upper cone for better view. \label{fig:orbits} }
\end{figure*}

In this section we show examples of the types of orbits discussed in the previous sections. The orbits are plotted in the cartesian coordinates $(X,Y,Z,W)$ defined in~\cite{anabalon11} as
\begin{eqnarray}
&&X=\sqrt{u+a^2}\sin\vartheta\cos\varphi \ , \, Y=\sqrt{u+a^2}\sin\vartheta\sin\varphi \ , \\
&&Z=\sqrt{u+a^2}\cos\vartheta\cos\psi \ , \, W=\sqrt{u+a^2}\cos\vartheta\sin\psi \ ,
\end{eqnarray}
where $r\in(0,\infty) \ , \vartheta \in (0, \frac{\pi}{2}) \ , \varphi \in (0, 2 \pi) \ , \psi \in (0, 2 \pi)$.
Since we cannot visualize a 4D picture we present 3D orbits for e.g. the coordinates $(X, Y, Z)$ in the Fig.\ref{fig:orbits}. Every 3D orbit is bounded by two cones with the opening angles defined by the roots of the polynomial $\Theta_\xi$~\eqref{xieom}. The spheres in the plots identify the outer and inner horizon.

Because of the bad choice of the coordinates the geodesics diverge at the horizons. This is associated with the directional change of the orbits in Fig.~\ref{fig:orbits}\subref{orb1},\subref{orb2_22},\subref{orb4} and~\subref{orb6}. It can be already seen in the differential equations~\eqref{dvarphidgamma} and~\eqref{dpsidgamma} because of the quadratic polynomial $\Delta$ which defines the horizons. This behaviour is known from the 4 dimensional Kerr(- de Sitter) space-time~\cite{Chandrasekhar83, Oneil, HLKK10}. In the next section~\ref{sec:observables} we proof that the frame-dragging effect is non zero in the 5-dimensional Myers-Perry space-time. This can be also seen in the MBO in the Fig.\ref{fig:orbits}\subref{orb1} and~\subref{orb6}, BO in the plots~\subref{orb2} and \subref{orb2_2} as well as in the escape orbit~\subref{orb5}. Here the orbit in the plot~\subref{orb2} has the energy value close to the energy of an unstable circular orbit in the potential well of the potential~\ref{pot6}.

Another interesting feature of the motion around rotating black holes can be at best seen in the pictures~\ref{orb2_22} and~\subref{orb6} (and also in the plots~\ref{orb1} and~\ref{orb4} but feebly seen). When the angular momentum (or both of them) of a test particle is negative in contrast to the positive angular momentum of the black hole, a test particle will be dragged in the direction of the black hole's rotation: as soon as it approaches the ergosphere it starts to corotate with the black hole. In the pictures~\ref{orb2_22} and~\subref{orb6} this behaviour correponds to the kinks shortly before the horizon (i.e. at the ergosphere which is not plotted here). That does not happen if one of the angular momenta has the same sign as the rotation parameter of the black hole (see e.g. picture~\ref{orb4}).

\begin{figure*}[th!]
\begin{center}
\subfigure[][BO. $r_0=1, a=0.4, K=10, E=8.07660455, \Phi=-4, \Psi=-0.5$. ]{\label{orb1_licht}\includegraphics[width=5.5cm]{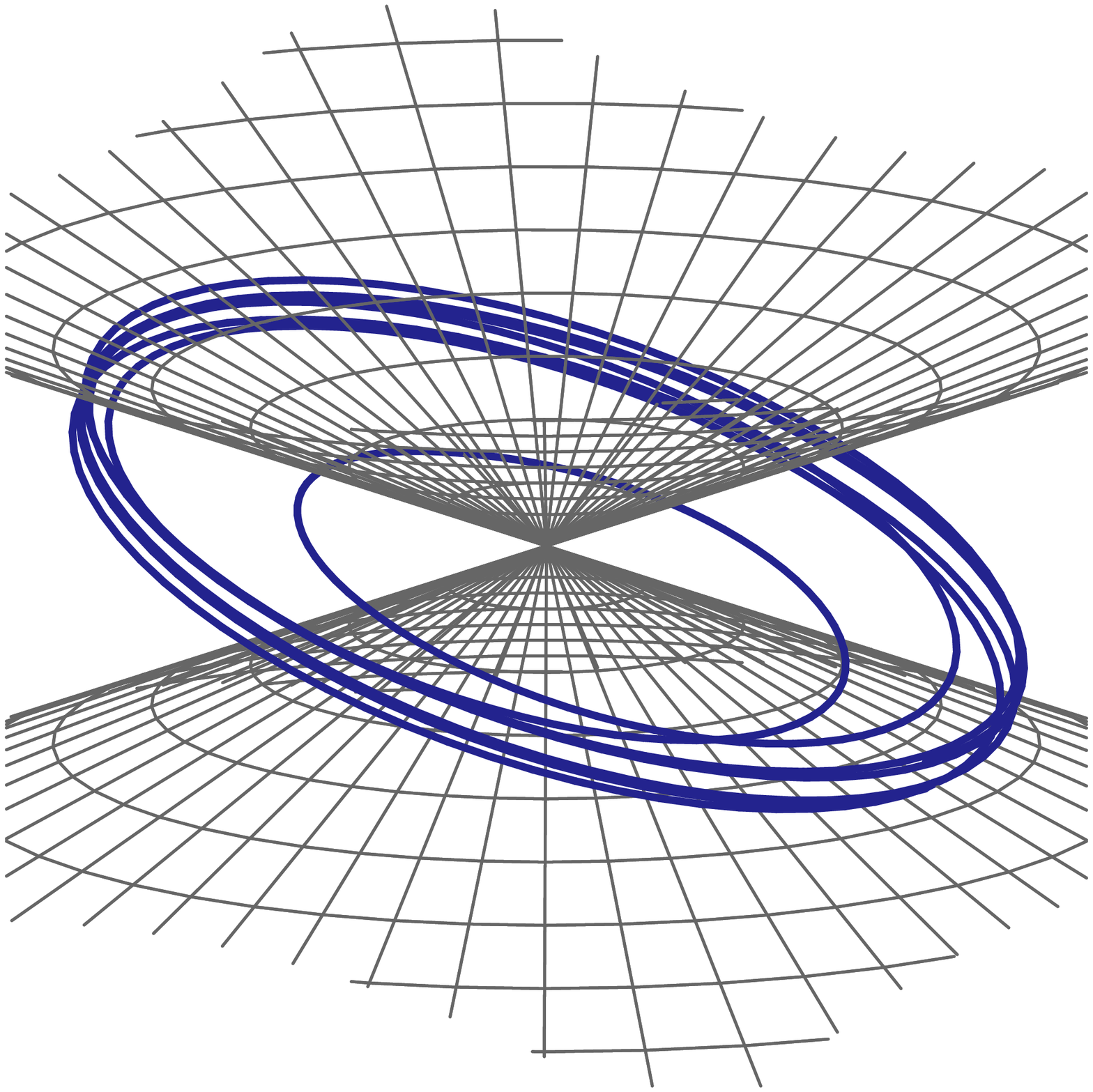}} \qquad 
\subfigure[][EO. $r_0=1, a=0.4, K=5, E=1.521633952, \Phi=-1, \Psi=-1$. ]{\label{orb2_licht}\includegraphics[width=5.0cm]{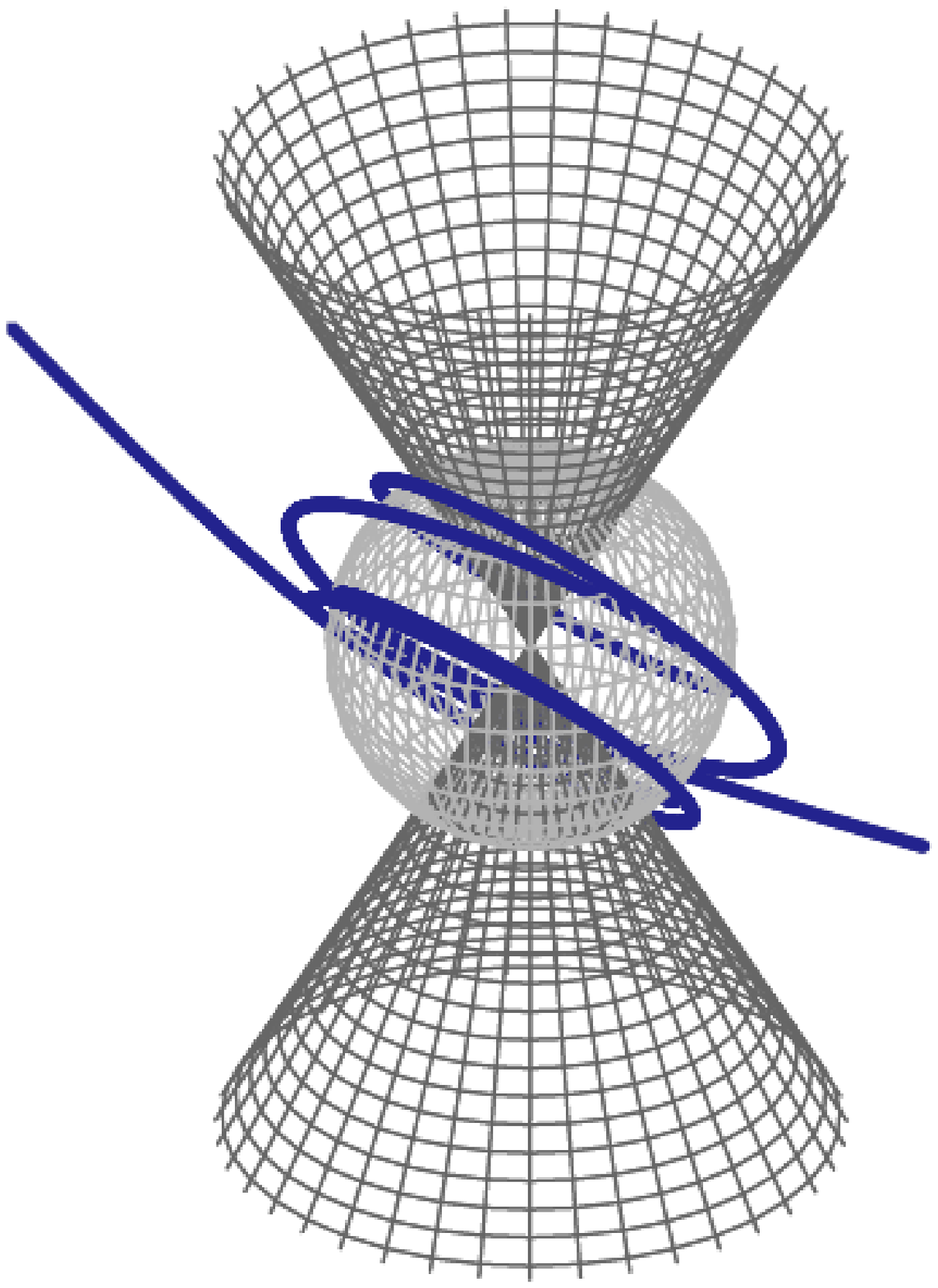}} \quad
\subfigure[][EO. $r_0=1, a=0.4, K=1, E=0.52, \Phi=0.01, \Psi=-0.01$. ]{\label{orb3_licht}\includegraphics[width=5.5cm]{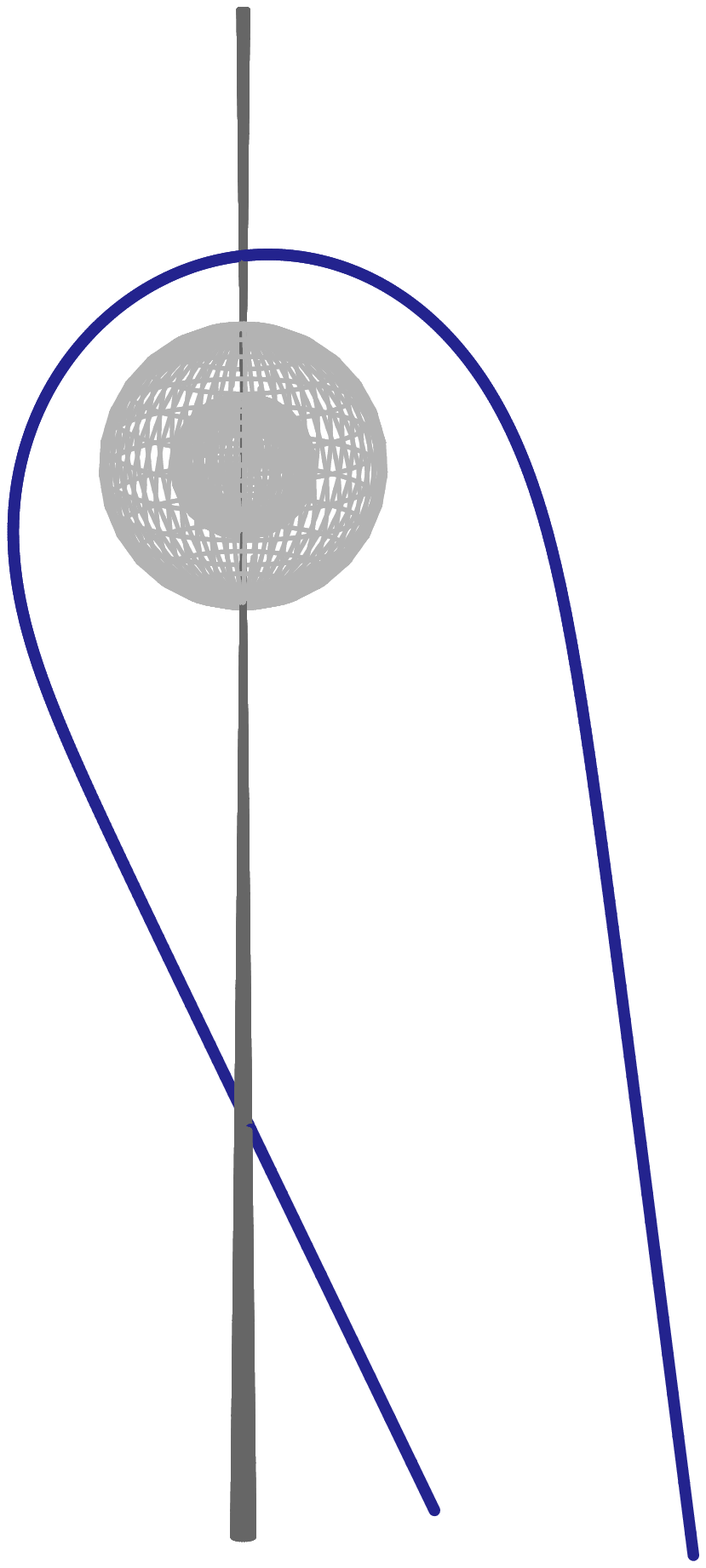}} 
\end{center}
\caption{Orbits for $\delta=0$ corresponding to the possible types specified in the potential plots~\ref{fig:potentiale_licht}. E.g., the depicted bound orbit in~\subref{orb1_licht} relates to the potential~\ref{pot6l}. Spheres have the radia of the horizons. \label{fig:orbits_licht} }
\end{figure*}

In the Fig.\ref{fig:orbits_licht} we show a bound and an escape orbit for massless particles illustrating similar properties as described above for the massive test particles.

\section{The observables}\label{sec:observables}

For observations one can define observables which do not depend on the chosen coordinate frame. These are e.g. the perihelion shift for bound orbits, 
the light deflection for escape orbits, the deflection angle for flyby orbits, or the Lense--Thirring effect which is present in axially symmetric space-times. 
We follow along the lines of~\cite{DrascoHughes04,FujitaHikida09} to calculate the observables. 

Consider the perihelion shift for a bound orbit BO or many world bound orbits MBO. 

The $u$-motion is periodic with period 
\begin{equation}
\omega_{u}= 2 \int_{u_{\rm min}}^{u_{\rm max}} \frac{du}{\sqrt{U}} = 2 \int_{e_1}^{e_{2}} \frac{dy}{\sqrt{P_3(y)}}  \ ,
\end{equation}
where $e_1$ and $e_{2}$ are the zeros of $P_3(y)$ related to $u_{\rm min}$ and $u_{\rm max}$. The corresponding orbital frequency is $\displaystyle{\frac{2\pi}{\omega_{u}}}$. 

The polar frequency is given by $\displaystyle{\frac{2\pi}{\omega_{\vartheta}}}$, where the polar period of the $\vartheta$--motion is defined as
\begin{equation}
\omega_{\vartheta}= 2 \int^{\vartheta_{\rm max}}_{\vartheta_{\rm min}} \frac{d\vartheta}{\sqrt{\Theta}} = - 2 \int^{\xi_{\rm max}}_{\xi_{\rm min}} \frac{d\xi}{\sqrt{\Theta_{\xi}}}
\end{equation}
for the angle $\vartheta_{\rm min}\leq\vartheta\leq\vartheta_{\rm max}$ where $\vartheta_{\rm min} \ , \vartheta_{\rm max} \in (0,\pi)$.

Next we calculate the secular accumulation rate of the time $t$:
\begin{equation}
\Gamma =  \frac{2}{\omega_{u}}  \int_{u_{\rm min}}^{u_{\rm max}}{ \left( (u+a^2) E + r_0^2\frac{\mathcal{E}}{\Delta} \right) \frac{du}{\sqrt{R}}}  
 = \frac{2}{\omega_{u}} \left( t_u(\gamma_{e_2}) - t_u(\gamma_{e_1}) \right) \ ,
\end{equation}
where $t_u$ is defined in~\eqref{t_u2} and $\gamma_{e_i}$ corresponds to the root $e_i$, $i=1,2$

The secular accumulation rates of the angles $\varphi$ and $\psi$ are given by: 
\begin{equation}
Y_{\varphi} = \frac{2}{\omega_{\vartheta}} \varphi_\xi \bigl|^{\xi_{\rm max}}_{\xi_{\rm min}} + \frac{2}{\omega_{u}} \varphi_u \Bigl|^{\gamma_{e_2} }_{\gamma_{e_1} }  
\end{equation}
and
\begin{equation}
Y_{\psi} = \frac{2}{\omega_{\vartheta}}  \psi_\xi \bigl|^{\xi_{\rm max}}_{\xi_{\rm min}} + \frac{2}{\omega_{u}} \psi_u \Bigl|^{\gamma_{e_2} }_{\gamma_{e_1} }  \ . 
\end{equation}

The orbital frequences $\Omega_u$, $\Omega_\vartheta$ and $\Omega_\varphi$, $\Omega_\psi$ are then given by:
\begin{equation}
\Omega_{u}=\frac{2\pi}{\omega_{u}}\frac{1}{\Gamma} \, , \qquad \Omega_\vartheta=\frac{2\pi}{\omega_{\vartheta}}\frac{1}{\Gamma} \, , \qquad \Omega_\varphi=\frac{Y_{\varphi}}{\Gamma} \, , \qquad \Omega_\psi=\frac{Y_{\psi}}{\Gamma} \ .
\end{equation}
The perihelion shift and the Lense--Thirring effects are defined as differences between these orbital frequences
\begin{align}
\begin{aligned}
\Delta_{\rm Perihel}^\varphi &= \Omega_\varphi - \Omega_u = \frac{1}{\Gamma} \left(Y_\varphi - \frac{2 \pi}{\omega_u} \right)\\[10pt]
\Delta_{\rm Perihel}^\psi &= \Omega_\psi - \Omega_u = \frac{1}{\Gamma} \left(Y_\psi - \frac{2 \pi}{\omega_u} \right)\\[10pt]
\Delta_{\rm Lense-Thirring}^\varphi &= \Omega_\varphi - \Omega_\vartheta = \frac{1}{\Gamma} \left(Y_\varphi - \frac{2\pi}{\omega_{\vartheta}} \right)\\[10pt]
\Delta_{\rm Lense-Thirring}^\psi &= \Omega_\psi - \Omega_\vartheta =  \frac{1}{\Gamma} \left(Y_\psi -  \frac{2\pi}{\omega_{\vartheta}} \right).
\end{aligned}
\end{align}

It seems to be not possible to measure a perihelion shift for a particle moving on an orbit located inside the inner horizon or for a many world bound orbit where only one part of the orbit can be seen and other parts are believed to happen in other universes. A deflection of light or massive particles by the gravitational field of the black hole seems to be more likely observable. This can be calculated in an analogous way (here in the $u$-motion the upper limit is infinity).

\section{Conclusions and Outlook}

In this paper we presented the analytical solution of the equations of motion in the 5D Myers-Perry space-time with equal rotation parameters. We integrated the encountered differentials of the first and third kind in terms of the Weierstra{ss}'s elliptic, zeta- and sigma-functions. We studied the properties of test particle motion defined by the $\vartheta$- and $u$- polynomials and completely characterized the possible types of motion. As examples we plotted the orbits of massive test particles for different values of the characterizing parameters.

The next step is to investigate the motion around a Myers-Perry black hole with non-equal rotation parameters~\cite{KaRe11}. It will elucidate the influence of the rotation of the black object on the motion of test particles further. It would be also interesting to go to higher dimensions, to add NUT parameters and to include a  cosmological constant. This will increase the order of the polynomials and the solution will be expressed in terms of hyperelliptic functions.

\section*{Acknowledgement}

We would like to thank Jutta Kunz and Claus Lämmerzahl for helpful discussions. V.K. acknowledges financial support of the German Research Foundation DFG. We also 
gratefully acknowledge support within the framework of the DFG Research Training Group 1620 {\it Models of gravity}.

\bibliographystyle{unsrt}

\end{document}